\documentclass[a4paper,11pt]{article}
\pdfoutput=1
\usepackage{jheppub}
\usepackage{amsmath,amssymb,euscript}
\usepackage{slashed}
\usepackage{xspace}
\usepackage{color}
\usepackage{accents}
\usepackage{hyperref}
\usepackage{epsfig}
\usepackage{xcolor}
\usepackage{xspace}
\usepackage{verbatim}
\usepackage{multirow}
\usepackage{booktabs,graphicx}
\usepackage{mathtools}
\usepackage{tabulary}
\usepackage{caption}
\usepackage{subcaption}
\usepackage{comment}
\usepackage{siunitx}
\usepackage{orcidlink}

\newcolumntype{K}[1]{>{\centering\arraybackslash}p{#1}}

\definecolor{cadmiumgreen}{rgb}{0.0, 0.42, 0.24}

\emergencystretch=\maxdimen
\begin{document}
\title{Tagging more quark jet flavours at FCC-ee at 91 GeV with a transformer-based neural network}

\author[1,2,4]{Freya Blekman \orcidlink{0000-0002-7366-7098},}
\author[3]{Florencia Canelli \orcidlink{0000-0001-6361-2117},}
\author[1]{Alexandre De Moor \orcidlink{0000-0001-5964-1935},} 
\author[1,3]{Kunal Gautam \orcidlink{0000-0002-1961-8711},}
\author[3]{Armin Ilg \orcidlink{0000-0001-9488-8095},}
\author[3]{Anna Macchiolo \orcidlink{0000-0003-0199-6957},}
\author[1,3]{Eduardo Ploerer \orcidlink{0000-0001-9336-4847},}

\affiliation[1]{Inter-university Institute for High Energies, Vrije Universiteit Brussel, 1050 Brussels, Belgium}
\affiliation[2]{Deutsches Elektronen-Synchrotron DESY, Notkestr. 85, 22607 Hamburg, Germany}

\affiliation[3]{Universit\"at Z\"urich, Winterthurerstr. 190, 8057 Z\"urich, Switzerland}
\affiliation[4]{Universit\"at Hamburg, Luruper Chaussee 149, 22761 Hamburg, Germany}

%\emailAdd{freya.blekman@desy.de}
%\emailAdd{armin.ilg@cern.ch}
\emailAdd{kunal.gautam@cern.ch}
\emailAdd{eduardo.ploerer@cern.ch}

		\begin{flushright}
		DESY/PUBDB-2024-01826\\
		\end{flushright}

\abstract{Jet flavour tagging is crucial in experimental high-energy physics. A tagging algorithm, \texttt{DeepJetTransformer}, is presented, which exploits a transformer-based neural network that is substantially faster to train than state-of-the-art graph neural networks.

The \texttt{DeepJetTransformer} algorithm uses information from particle flow-style objects and secondary vertex reconstruction for $b$- and $c$-jet identification, supplemented by additional information that is not always included in tagging algorithms at the LHC, such as reconstructed $K_{S}^{0}$ and $\Lambda^{0}$ and $K^{\pm}/\pi^{\pm}$ discrimination. The model is trained as a multiclassifier to identify all quark flavours separately and performs excellently in identifying $b$- and $c$-jets. 
An $s$-tagging efficiency of $40\%$ can be achieved with a $10\%$ $ud$-jet background efficiency. The performance improvement achieved by including $K_{S}^{0}$ and $\Lambda^{0}$ reconstruction and $K^{\pm}/\pi^{\pm}$ discrimination is presented.

The algorithm is applied on exclusive $Z \to q\bar{q}$ samples to examine the physics potential and is shown to isolate $Z \to s\bar{s}$ events. Assuming all non-$Z \to q\bar{q}$ backgrounds can be efficiently rejected, a $5\sigma$ discovery significance for $Z \to s\bar{s}$ can be achieved with an integrated luminosity of $60~\text{nb}^{-1}$ of $e^{+}e^{-}$ collisions at $\sqrt{s}=91.2~\mathrm{GeV}$, corresponding to less than a second of the FCC-ee run plan at the $Z$ boson resonance.}

\maketitle
\flushbottom

\clearpage
%---------------------------------------------------------------------------

\section{Introduction}
\label{sec:intro}
The Standard Model (SM) of particle physics \cite{Glashow:1959wxa,Salam:1959zz,Weinberg1967,tHooft1972} is one of the most successful scientific theories describing the fundamental particles and their interactions. The last piece of this model, the Higgs boson, was discovered \cite{20121, 201230} at the Large Hadron Collider (LHC) \cite{Bruning:782076} in $2012$, and the precise study of its properties will remain mostly superficial at the LHC due to high irreducible backgrounds from other SM processes while isolating Higgs boson events.

One of the main motivations for proposed future lepton colliders \cite{Benedikt:2651299, Adolphsen:1601969, CEPCStudyGroup:2018ghi, Linssen:2012hp} is the precise measurement of SM parameters, like precision studies of the hadronic decay of the $Z$ boson and greatly improved sensitivity to the couplings of the Higgs boson to the bottom ($b$) and charm ($c$) quarks and gluons ($g$) \cite{European:2720131, de_Blas_2020, dEnterria:2017dac}. 
Achieving these objectives requires an efficient reconstruction and identification of the hadronic decays of these particles. The feasibility of studying the decay of the Higgs boson to the strange ($s$), up ($u$), and down ($d$) quarks depends on the collider and detector performance and is currently under investigation in the field. 
It is well established that efficient and accurate jet flavour identification is essential to exploit the maximal physics potential of future collider experiments \cite{Azzi2021, An_2019, Asner:2013psa, Albert:2022mpk, Abramowicz_2017}.

The state-of-the art is shortly reviewed in the rest of Section \ref{sec:intro}. Section \ref{sec:experiment} summarises the FCC-ee collider, the IDEA detector concept, and the used simulated samples and provides minimal event selection requirements. Section \ref{sec:vtxreco} briefly describes the algorithms used to reconstruct displaced decay vertices and their performance. Section \ref{sec:NN} introduces the attention mechanism and Transformer models and outlines the description of the input features and the network architecture used for tagging. Finally, the obtained results and the performance of the flavour-tagging algorithm in $Z$ boson signatures are presented in Section \ref{sec:performance} and \ref{sec:Zss}, respectively.

\subsection{Review of Jet Flavour Tagging} \label{subsec:relatedwork}
Jets originating from the $b$ and $c$ quarks contain hadrons with significant lifetimes that travel distances of the order of millimeters from the interaction point before decaying into lighter hadrons. 
The heavy flavour tagging algorithms used at the Large Electron-Positron collider (LEP) \cite{2004,Proriol:1950599} and the Tevatron \cite{Abazov_2010,Freeman_2013} experiments exploited variables derived from the displaced charged tracks originating from these decayed $b$ or $c$ hadrons to distinguish the heavy flavoured jets from $s$, $u$, $d$ quark and gluon jets. These charged tracks are commonly clustered to reconstruct the original decay vertices of the $b$ and $c$ hadrons, also called secondary vertices (SVs). The properties of these SVs, like their mass and displacement, are some of the most important inputs used to identify $b$- and $c$-jets.

The understanding and performance of jet flavour tagging at the LHC has steadily been improving and heavily relies on machine learning (ML) \cite{atlas_run2_ftagging,CMS:2017wtu}, which also inspires flavour tagging algorithms for the FCC-ee \cite{Qu:2019gqs, Bols_2020}.

ML approaches are uniquely suited to classify jet flavours, where training samples are abundant in the form of Monte Carlo (MC) simulation. Still, the underlying dynamics of jet formation and hadronisation are not always well understood. With the advent of ML techniques, including Neural Networks (NNs) and Boosted Decision Trees (BDTs), approaches relying on single physics-motivated variables for jet flavour discrimination were significantly outperformed \cite{CMS:2017wtu, ATLAS:2015thz, ATLAS:2019bwq, Mondal_2024}. Since then, a multitude of architectures and jet representations have found success in discriminating jet flavours, including Dense Neural Networks (DNNs) \cite{Luo:2017ncs}, Recurrent Neural Networks (RNNs) \cite{Guest:2016iqz}, Convolutional Neural Networks (CNNs) \cite{Komiske:2016rsd, ATLAS:2017dfg}, and Graph Neural Networks (GNNs) \cite{Qu:2019gqs, Mikuni:2020wpr, ATLAS:2022rkn}.

Among the most successful of these are Graph-based architectures such as \texttt{ParticleNet} \cite{Qu:2019gqs} that represent jets as sets of nodes (jet constituents) and edges (some pairwise defined feature, often the difference in a given variable of jet constituents). In particular, networks combining a self-attention mechanism \cite{vaswani2023attentionneed} to exploit the relative importance of constructed features, dubbed Transformer Networks, have achieved state-of-the-art performance in the task of jet flavour tagging \cite{Mondal_2024, duperrin2023flavourtagginggraphneural, ATLAS:2023ber, Qu:2022mxj}. Particle Transformer (ParT) \cite{Qu:2022mxj} combines a graph representation of jets with an attention mechanism.
In this work, a pure Transformer architecture, \texttt{DeepJetTransformer}, similar to the ParT (plain) variant introduced in Ref. \cite{Qu:2022mxj}, is presented for the task of jet flavour identification at future lepton colliders, using the FCC-ee with the IDEA detector concept as a benchmark \cite{Benedikt:2651299, Bedeschi:2021bS}. \texttt{DeepJetTransformer} is relatively lightweight and requires much less computational time compared to graph-based architectures \cite{Bedeschi_2022, Qu:2022mxj, 10363595}, yet achieves comparable tagging performance.

\subsection{The \textit{Z} boson at the FCC-ee} \label{subsec:ZatFCCee}
After the discovery of the $Z$ boson at the Super Proton Synchrotron (SPS) at CERN in $1983$ \cite{UA1:1983mne,UA2:1983mlz}, this neutral vector boson was extensively studied at the LEP collider and the SLAC Linear Collider. The existence of the $Z$ boson confirmed the electroweak mixing \cite{Woods:298820,Dam:348463} and the measurement of its width constrained the number of neutrino generations to three \cite{Barate:396345,DELPHI:1989cmk,OPAL:1989hoi,PhysRevLett.63.724,2006}.

The proposed FCC-ee program provides a unique opportunity to push the $Z$ boson measurements to their ultimate limit. The four-year-long FCC-ee run at and around the $Z$ resonance will produce an unprecedented $6\times10^{12}$ total decays. The integrated luminosity expected at the $Z$ resonance at FCC-ee is $125~\text{ab}^{-1}$, about $10^6$ times that of LEP. The statistical errors on the mass and width of the $Z$ boson can be reduced from $1.2$ MeV and $2$ MeV to $5$ KeV and $8$ KeV \cite{Benedikt:2651299}, respectively. Lower center-of-mass energy spread due to beam energy calibration will benefit in reducing the systematic uncertainty of these quantities. Measuring the forward-backward and polarisation asymmetries is a powerful method to estimate the effective weak mixing angle, $\sin^2\theta^{\text{eff}}_{\text{W}}$, for which the statistical uncertainty is expected to reduce to about $10^{-6}$, corresponding a more than thirty-fold improvement \cite{Benedikt:2651299}.

Studying the hadronic decay channels of the $Z$ boson is a very important aspect of the FCC-ee physics program. The couplings and decay widths of the $Z$ boson have only been measured to the heavier quarks, $b$ and $c$. The only study of the $s$ quark decay of the $Z$ boson available in the literature is preliminary \cite{Boudinov:1998fao}. For the lighter quarks, $s$, $u$, and $d$, these properties are typically only listed collectively for up-type and down-type quarks \cite{Workman:2022ynf}. Similarly, the axial and vector couplings have also been collectively measured for up-type and down-type quarks \cite{Workman:2022ynf}.

Future colliders with a dedicated $Z$ boson run, like FCC-ee, will improve the precision of all these measurements and make the $s$ quark, and potentially the $u$, and $d$ quarks, accessible. Individual measurements of the quark vector and axial couplings should be possible via their forward-backward asymmetries, corresponding partial decay widths of the $Z$ boson, and the precise knowledge of $A_e$, the asymmetry parameter of the $e^-e^+$ pair. The experimental systematic uncertainties corresponding to these measurements are also expected to drastically improve due to better detector designs \cite{Benedikt:2651299}.

\subsection{Strange Jet Tagging} \label{subsec:stagingintro}
The discrimination of $s$-jets is widely regarded as one of the most challenging types of jet discrimination. Thus, it has received considerably less attention than its heavy-flavour counterparts, or indeed gluon discrimination. At the core of the problem is the fact that unlike in the discrimination of quarks vs gluons, which relies heavily on properties following from their differing colour factors $C_{F} = 4/3$ vs $C_{A} = 3$, or heavy flavour tagging, which relies on displaced vertices of $b/c$ hadrons, strange quarks are treated identically to down quarks by QCD and Electroweak theory in the massless limit prior to their decay. Discriminating strange and down jets is particularly challenging due to the same fractional charge of the initiating quarks. In practice, however, strange hadrons carry a larger fraction of the total scalar momentum of strange jets, compared to hadrons consisting of up and down ($ud$) quarks. The total scalar momentum is obtained by summing over the scalar momentum of all jet constituents. This idea was also explored in the context of hadron colliders \cite{Nakai:2020kuu}. Strange jets tend to have a higher kaon multiplicity and a lower number of pions than $u$- and $d$-jets. Therefore distinguishing $K^{\pm}$ and $\pi^{\pm}$ and reconstructing $K^{0}_{S}$ is crucial for strange jet identification \cite{Erdmann_2020,Nakai:2020kuu, Erdmann:2020ovh}.

SLD \cite{SLD:1984aa} tagged $Z \to s\bar{s}$ events by looking for the absence of reconstructed $b$ and $c$ hadrons and the presence of $K^{\pm}$ or $K^{0}_{S}$ \cite{PhysRevLett.85.5059}. Particle identification (PID) was performed at SLD, as at DELPHI \cite{delphi_loi}, with a RICH detector. At most other detectors, energy loss ($dE/dx$) was used for PID \cite{opal_loi, l3_loi}, with the addition of timing at ALEPH \cite{aleph_loi}. The detector concepts at the FCC-ee foresee the use of techniques like energy loss ($dE/dx$) \cite{Lippmann2012}, ionisation cluster counting ($dN/dx$) \cite{Cataldi:1996mz}, time-of-flight \cite{Klempt:1999tx}, and compact-Ring Imaging CHerenkov (RICH) detectors.

Tagging strange jets at future colliders has been explored as a probe to perform precision measurements in the Higgs sector \cite{DuarteCampderros2020,Albert:2022mpk}, and the impact of using $dN/dx$ and time-of-flight on strange tagging performance for jets originating from Higgs boson decay was studied using a graph neural network \cite{Bedeschi_2022}. In this work, \texttt{DeepJetTransformer} is used to isolate $Z \to s\bar{s}$ events from the exclusive hadronic decays of the $Z$ boson in the FCC-ee environment. The excess momentum carried by strange hadrons is exploited, firstly by including V$^0$ variables and secondly through $K^{\pm}/\pi^{\pm}$ discrimination. The cleaner environment at lepton colliders and the powerful PID capabilities of the proposed detectors facilitate making strange jet tagging feasible.

\section{Experimental Environment}
\label{sec:experiment}

\subsection{FCC-ee}
\label{subsec:fccee}
The Future Circular Collider (FCC) integrated project \cite{FCC:2018byv,fccfs-midterm} aims to build $e^+e^-$, $pp$, and $ep$ colliders in a $90.7$ km circular tunnel in the Geneva region. FCC-ee \cite{Benedikt:2651299} is a proposed $e^+e^-$ collider and the first stage of the FCC integrated project. It is currently planned to run at four different center-of-mass energy modes, starting from around $91.2$ GeV at the Z-pole to $365$ GeV, over the $t\bar{t}$ threshold. The unprecedented luminosities at the FCC-ee uniquely facilitate tests of the SM and, at the same time, present novel challenges in reducing systematic errors. The circular collider design provides the opportunity for four interaction points, each of which can host a different detector design. Such detector concepts \cite{Bedeschi:2021bS, https://doi.org/10.48550/arxiv.1911.12230, Francois2022} are currently being studied, of which the IDEA detector concept \cite{Bedeschi:2021bS} has been used in this study.

\subsection{IDEA Detector Concept}
\label{subsec:idea}
A fast simulation of the IDEA detector concept \cite{idea_delphes} has been implemented in \texttt{Delphes} \cite{de_Favereau_2014} and used for the simulation of the samples used in this work. A spherical coordinate system is used with its origin at the center of the detector system and the positive $z$ axis in the direction of travel of the incoming electron. The polar angle, $\theta$, is defined as the angle between the radial line and the positive $z$ axis and the azimuthal angle, $\phi$, is defined as the angle of rotation of the radial line around the positive $z$ axis.

The innermost part of the IDEA detector is the monolithic active pixel sensor (MAPS) based vertex detector, which consists of three inner layers with a space point resolution of \SI{3}{\micro\meter}, and two outer barrel and three disk layers on each side with a space point resolution of \SI{7}{\micro\meter}. The innermost layer is positioned at a radius of $1.7$ cm. The vertex detector is enclosed by the drift chamber incorporating $112$ layers of \SI{100}{\micro\meter} resolution. The multiple scattering of particles is minimal thanks to the main gas component being Helium. Two layers of silicon sensors surround the drift chamber to provide a very precise space point measurement. A single-hit resolution of $\SI{7}{\micro\meter}$ ($\SI{90}{\micro\meter}$) along $\phi$ ($z$) is assumed. These sit inside a solenoid magnet with a $2$ T magnetic field. It is followed by a dual-readout calorimeter that is sensitive to independent signals from the scintillation and the Cerenkov light production. This results in a good energy resolution for both electromagnetic and hadronic showers. The calorimeter is enveloped by the muon system consisting of layers of chambers embedded in the magnet return yoke. The detector geometry has been modified since generating the event samples, and further optimisation is in progress.

\subsection{Event Samples and Jet Reconstruction} \label{subsec:sim}

The simulated event samples used for training and evaluation consist of the process $e^{+}e^{-} \rightarrow Z \rightarrow q\bar{q}$, where $q \equiv b, c, (u, d, s)$, at the center-of-mass energy ($\sqrt{s}$) of $91.2$ GeV. \texttt{Pythia8.303} \cite{https://doi.org/10.48550/arxiv.2203.11601} is used for event generation, parton showering, and hadronisation. \texttt{Delphes} \cite{de_Favereau_2014} is used for event reconstruction assuming the IDEA detector concept \cite{Bedeschi:2021bS,idea_delphes}. A tracking efficiency of $99.7\%$ is assumed for electrons, muons and charged hadrons with 3-momentum magnitude $|p| > 0.5$ GeV that lie within acceptance. This efficiency is reduced to $65\%$ ($4\%$) for $0.5 > |p| > 0.3$ GeV ($|p| < 0.3$ GeV). Fake tracks are not considered.

Jet clustering is performed on the particle flow-style objects reconstructed by \texttt{Delphes} with \texttt{FastJet-3.3.4} \cite{Cacciari_2012} using the exclusive $e^{+}e^{-}~k_{\text{T}}$ algorithm \cite{Catani:1991hj}. Other jet clustering algorithms like the anti-$k_{\text{T}}$ algorithm \cite{Cacciari_2008} and the generalised $e^+e^-~k_{\text{T}}$, also referred to as the inclusive $e^+e^-~k_{\text{T}}$, algorithm \cite{Cacciari_2012} were also considered. The exclusive $e^{+}e^{-}~k_{\text{T}}$ algorithm, which creates irregularly shaped jets and, in this study, requires exactly two jets, is very robust against gluon emissions and gluon splitting. Since the exclusive $e^+e^-~k_{\text{T}}$ algorithm clustered jets include all reconstructed final particles, they were observed to satisfy the requirements of this study by most accurately reproducing the $Z$ boson reconstructed invariant mass signature. No additional selections were applied to the samples for training and evaluation of the jet flavour tagger.

In this study, the jets are assigned an MC flavour as the flavour of the quarks to which the $Z$ boson decays. Besides simplicity, this has the added benefit that other studies for future facilities use the same definition.

A separate set of event samples was generated with the process $e^{+}e^{-} \rightarrow Z(\rightarrow \nu\nu)H \rightarrow q\bar{q}$, where $q \equiv b, c, (u, d, s)$, at $\sqrt{s}$ of $240$ GeV. The same reconstruction and jet clustering were applied as for the samples at the $Z$ resonance. Training and evaluation of \texttt{DeepJetTransformer} were performed with these samples for comparison with other taggers.

\section{Vertex Reconstruction} \label{sec:vtxreco}
Vertex reconstruction is essential to find the primary interaction vertex and the secondary decay vertices of the long-lived $b$, $c$, and $s$ hadrons. It helps improve the $b$- and $c$-tagging performance and aids in $s$-tagging. Charged tracks can be fitted to reconstruct the primary and the displaced secondary vertices. These displaced vertices can either be the decay vertices of $b$ and $c$ hadrons (SVs) or those of the long-lived hadrons containing $s$ quarks, like $K^{0}_{S}$ or $\Lambda^{0}$, commonly referred to as V$^0$s, which are particles that decay into a pair of oppositely charged tracks. All displaced vertices except V$^0$s are referred to as SVs. The properties of SVs and V$^0$s, such as their masses, displacements, and charged track multiplicities, can be used to identify the decaying hadrons and, in effect, the jet flavour. The SVs can even be used to reconstruct the entire hadronic decay chain. Similarly, reconstructing and identifying the V$^0$ vertices can be used to identify $s$-jets, as $K^{0}_{S}$ and $\Lambda^{0}$ are the particles carrying most of the momentum of some $s$-jets \cite{Albert:2022mpk}. Distinguishing V$^0$s from SVs also helps to reduce the misidentification of some $b$- and $c$-jets as $s$-jets.

The vertex reconstruction in this study has been performed using an implementation of the vertexing module of the \texttt{LCFIPlus} framework \cite{Suehara_2016, Gautam:2839048}. It has been implemented in \texttt{FCCAnalyses} \cite{fccana}, the FCC software framework, using a $\chi^{2}$-based vertex fitter \cite{vertex_fitter}. The constraints and parameters have been kept the same as in Ref. \cite{Suehara_2016}. The algorithm first identifies the tracks forming V$^0$s. Unlike in standard vertex reconstruction algorithms, the V$^0$s are not discarded but stored and assigned a particle ID based on the set of constraints that they pass, summarised in Table \ref{tab:v0constraints}. The tracks originating from the primary vertex or V$^0$ candidates are not considered while reconstructing SVs.

The properties of the SVs and V$^0$s, along with more variables, are used as input to train the neural network tagger described in Section \ref{sec:NN}.

\begin{figure}%[b]
    \centering
    \begin{subfigure}[b]{0.495\textwidth}
        \centering
        \includegraphics[width=\textwidth]{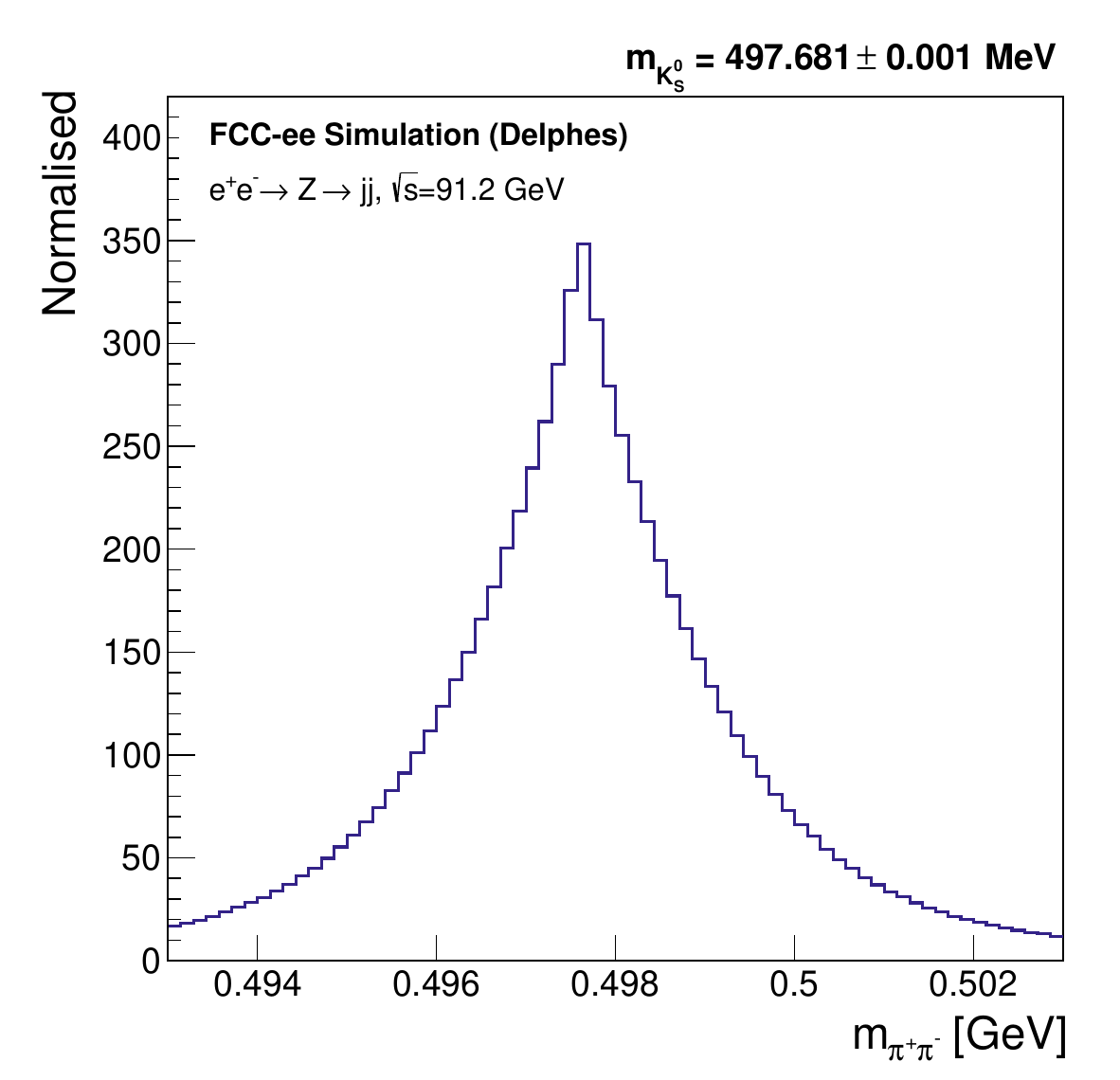}
        \caption{Reconstructed $K_{S}^{0}$}
        \label{fig:Ks}
    \end{subfigure}
    \hfill
    \begin{subfigure}[b]{0.495\textwidth}
        \centering
        \includegraphics[width=\textwidth]{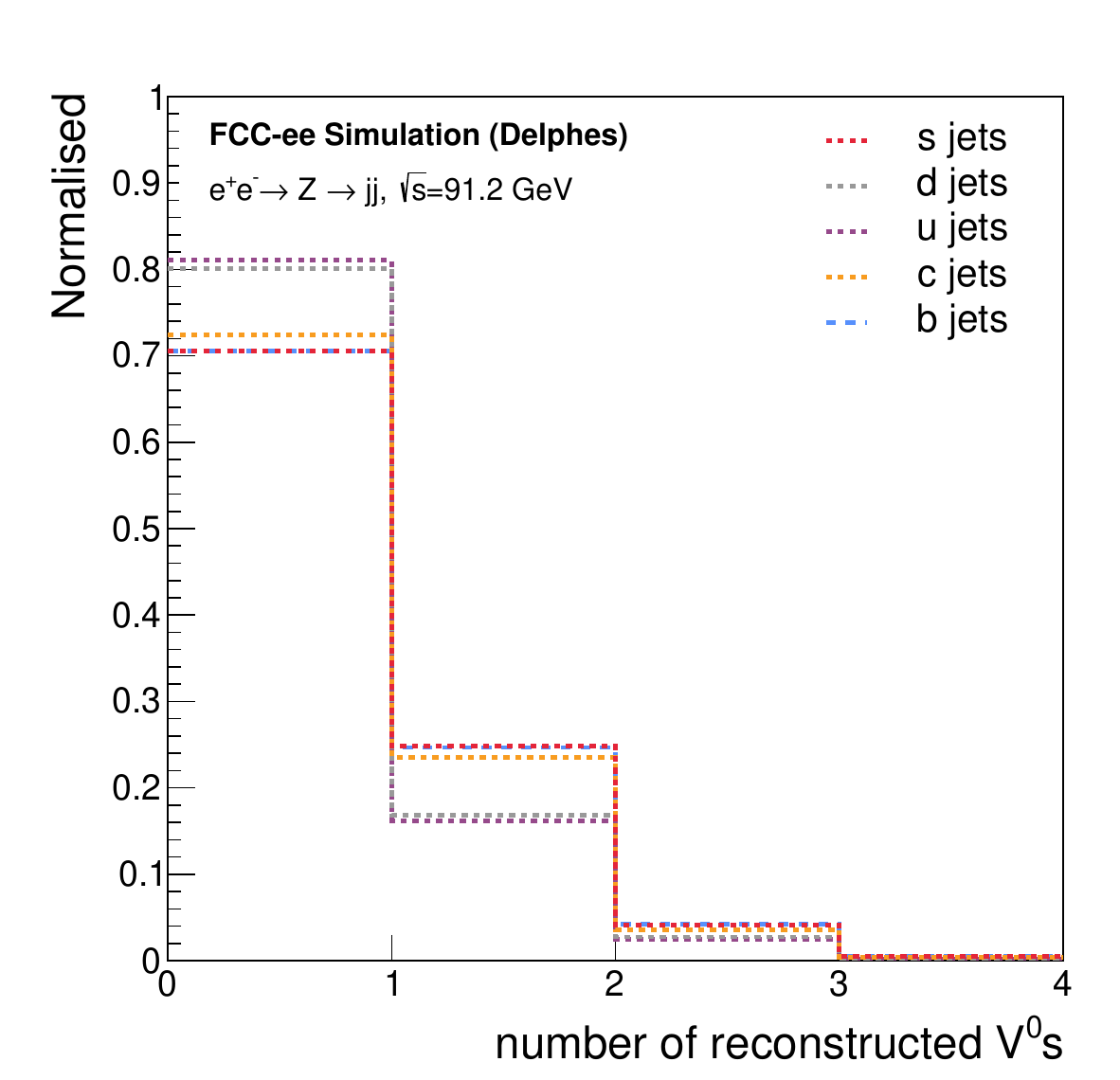}
        \caption{V$^0$ multiplicities}
        \label{fig:n_V0}
    \end{subfigure}
    \caption{Performance of V$^0$ reconstruction. (a) Invariant mass distribution of reconstructed $K_{S}^{0}$ vertices. The quoted mass is the mean and the error on the mean of the distribution. (b) The reconstructed V$^0$ multiplicity in jets from $e^{+}e^{-} \rightarrow Z\rightarrow q\bar{q}$ events at $\sqrt{s}=91.2$ GeV, where $q \equiv u,d,s,c,b$. The distributions for $b$- and $s$-jets overlap almost perfectly.}
    \label{fig:V01}
\end{figure}

\subsection{V\texorpdfstring{\raisebox{0.5ex}{\scriptsize 0}}{0} Vertex Reconstruction} \label{subsec:V0}
Two processes are considered: $K_{S}^{0} \to \pi^{+}\pi^{-}$ and $\Lambda^{0} \to p\pi^{-}$. The invariant mass of the reconstructed $K_{S}^{0}$s can be seen in Figure \ref{fig:Ks}, demonstrating a good reconstruction of V$^0$s and their properties. The mass of the tracks used to calculate the invariant mass of the V$^0$ is decided based on the set of constraints the V$^0$ passes with a certain permutation of the two tracks. In contrast, all tracks are assumed to be pions in the invariant mass calculation for the SVs.

\begin{table}
\centering
\begin{tabular}{l|c c c c}
     & \multicolumn{2}{c}{$K_{S}^{0}$} & \multicolumn{2}{c}{$\Lambda^{0}$} \\
     & tight & loose & tight & loose \\
    \hline
    \hline
    M [GeV] & $[0.493, 0.503]$ & $[0.488, 0.508]$ & $[1.111, 1.121]$ & $[1.106, 1.126]$ \\
    r [mm] & $>0.5$ & $>0.3$ & $>0.5$ & $>0.3$ \\
    $\hat{\text{p}}\cdot\hat{\text{r}}$ & $>0.999$ & $>0.999$ & $>0.99995$ & $>0.999$
\end{tabular}
\caption{Summary of the default V$^0$ selection criteria \cite{Suehara_2016}. M is the invariant mass, and p is the momentum of the V$^0$ candidate. r is the distance of the V$^0$ candidate from the primary vertex. The collinearity of the V$^0$ candidate is defined as $\hat{\text{p}}\cdot\hat{\text{r}}$. The set of `tight' constraints has been used to identify V$^0$s in this study, while the set of `loose' constraints has been used to remove the V$^0$ background while reconstructing SVs.}
\label{tab:v0constraints}
\end{table}

Figure \ref{fig:n_V0} displays the V$^0$ multiplicity in jets from $Z\to q\bar{q}$ events. No reconstructed V$^0$s are found for most of the jets. But, a higher fraction of heavy- and strange-flavoured jets contain reconstructed V$^0$s than $u$- and $d$-jets, which justifies the importance of V$^0$ rejection before attempting to reconstruct SVs. It is also evident that more $s$-jets have one or more reconstructed V$^0$s than $u$- and $d$-jets, making V$^0$s an important discriminator of $s$-jets against lighter quark jets.

\subsection{Secondary Vertex Reconstruction} \label{subsec:SV}
\begin{figure}
    \centering
    \includegraphics[width=\textwidth]{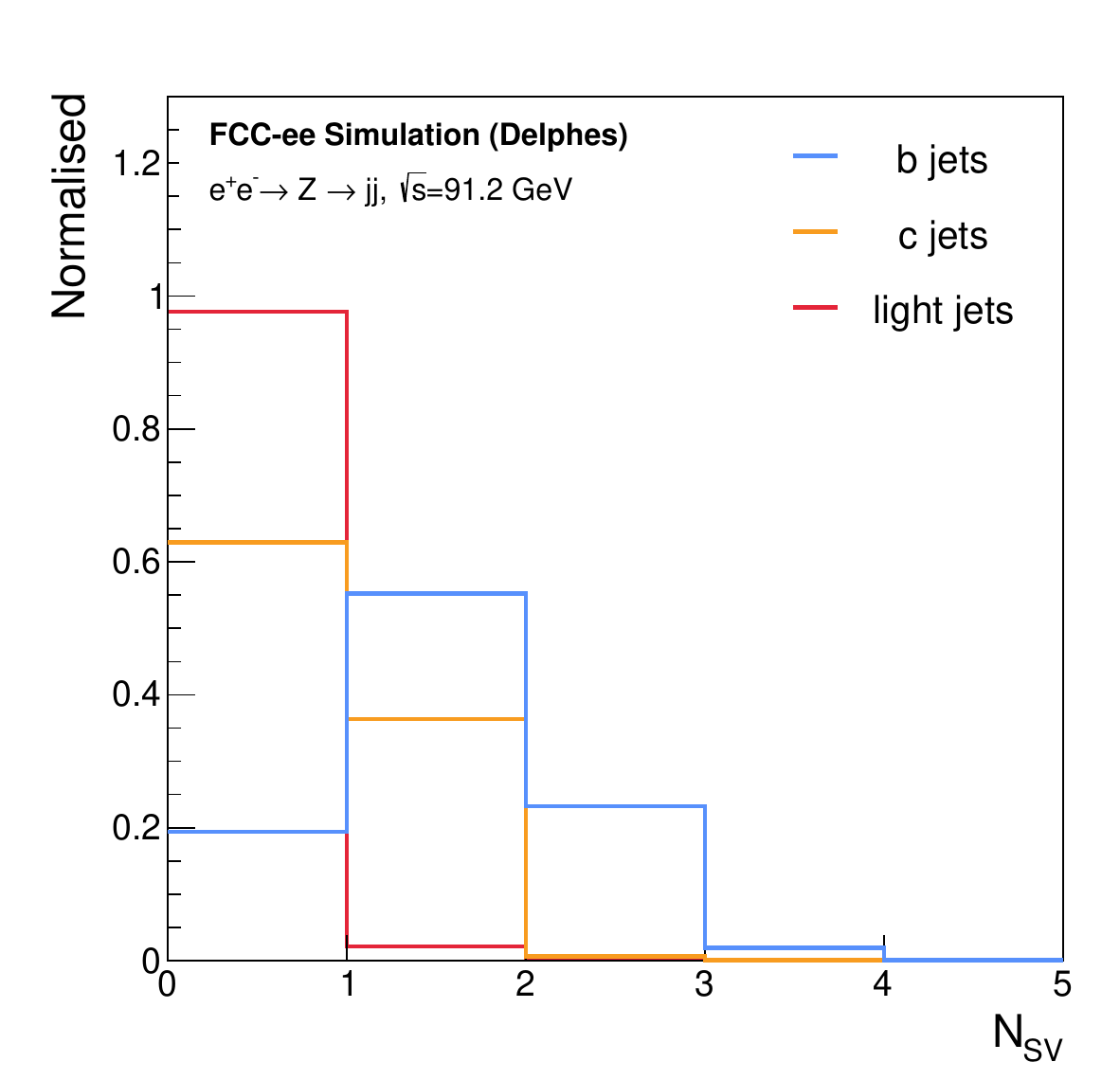}
    \caption{SV multiplicity in jets from $e^{+}e^{-} \rightarrow Z \rightarrow q\bar{q}$ events. The term ``light jets" here collectively refers to $u$-, $d$-, and $s$-jets.}
    \label{fig:SV_n}
\end{figure}

Due to the near-diagonal CKM matrix, the cascading decay chain of heavier quarks is expected to be $b \to c \to s \to (u,d)$. Hence, the SV multiplicity tends to be higher in $b$-jets compared to $c$-, $s$-, $u$-, and $d$-jets, as shown in Figure \ref{fig:SV_n}.

\section{DeepJetTransformer} \label{sec:NN}

Since the introduction of \texttt{ParticleNet} \cite{Qu:2019gqs}, the concept of a Particle Cloud has become the prevailing representation of jet structure. A Particle Cloud considers the jet as an unordered set of jet constituents of varying length. Elements of differing nature, such as charged, neutral particles, or SVs associated with the jet, are considered to create the most complete and accurate representation. This representation concept was used to build the presented model, the key element of which, the unordered set of particles, requires the construction of a model invariant under the permutation of the jet constituents\footnote{Permutation invariance is in opposition to most Transformer models established around the principle of causality \cite{vaswani2023attentionneed, devlin2019bert}.}, a benefit also used by other transformer-based taggers \cite{Qu:2022mxj}. Moreover, Transformers possess the essential property of full connectivity between jet constituents via the attention mechanism \cite{vaswani2023attentionneed}. This enables the model to capture subtle correlations among jet constituents, enhancing the high-level features used for jet discrimination.

A structure based on Transformer blocks was thus chosen for this study. Previous research has indicated that Transformer models offer enhanced performance and increased efficiency, particularly compared to graph models \cite{Qu:2022mxj, 10363595}. The subsequent sections will elaborate on the inputs to the neural network and the fundamental characteristics of Transformer models and provide a detailed description of the specific model, \texttt{DeepJetTransformer}, which has been developed for this study.

\subsection{Input Features} \label{subsec:input}
The properties of each jet and its constituents represent different categories of input features available for model training. All input features are built using information reconstructed with \texttt{Delphes} detector simulation unless stated otherwise. The jet kinematics are represented by variables defined using its $4-$momentum, as detailed in Table \ref{tab:jetvars}. Many future collider detector concepts are designed to be used with a particle flow algorithm \cite{BUSKULIC1995481, CMS-PAS-PFT-09-001}. Therefore, jet constituents are subdivided into five sets according to the typical particle flow candidate categories:  charged hadrons, neutral hadrons, electrons and positrons $(e^{\pm})$, photons $(\gamma)$, and muons $(\mu^{\pm})$. Kinematic variables are defined for each jet constituent using its $4$-momentum, as listed in Table \ref{tab:jetconstvars}. For each jet up to 25 charged jet constituents and 25 neutral jet constituents are considered. This is enforced by truncating the input feature array of a given jet if the number of charged/neutral jet constituents is more than 25. Conversely, if the number of charged/neutral jet constituents is less than 25, then the input feature array is zero-padded. 

Charged tracks are first fitted to find the V$^0$s and the remaining tracks are used to reconstruct SVs. Feature variables are defined separately for both classes of reconstructed vertices (V$^{0}$s and SVs) and are listed in Table \ref{tab:svvars}. Up to 4 V$^0$s and 4 SVs are considered per jet. The V$^0$ and SV input feature arrays are likewise truncated/zero-padded. The distinguishing power of some of these variables is discussed below.

\begin{table}[ht]
\centering
    \begin{tabular}{c|c}
      Input Feature & Description \\
      \hline
      \hline
       $|p|,~E,~m$  & 3-momentum magnitude, energy, and invariant mass of the jet \\
       $\theta,\phi$ & polar and azimuthal angle of the jet axis \\
       $\text{N}_{\text{charged}}$ & charged particle (track) multiplicity in the jet \\
       $\text{N}_{\text{neutral}}$ & neutral particle multiplicity in the jet \\
       & jet angularity \cite{Larkoski:2014pca} as sum of normalized jet constituent energy ($z_{i}$)\\
       $\lambda^{\kappa}_{\beta} = \Sigma_{i\in \text{jet}} z^{\kappa}_{i}R^{\beta}_{i}$ & and angular distance to jet axis ($R_{i}$) for ($\kappa = 0$, $\beta = 0$), \\
       & ($\kappa = 1$, $\beta = 0.5$), ($\kappa = 1$, $\beta = 1$), ($\kappa = 1$, $\beta = 2$), ($\kappa = 0$, $\beta = 2$)\\
       \hline
       \texttt{isU/D/S/C/B} & MC flavour assigned to the jet \\
    \end{tabular}
\caption{\label{tab:jetvars}Description of global features associated with each jet}
\end{table}

\begin{table}[ht]
\centering
    \begin{tabular}{c|c}
      Input Feature & Description \\
      \hline
      \hline
       $D_{0} (z_{0})$  & signed transverse (longitudinal) impact parameter \\
       $D_{0}/\sigma_{D_{0}} (z_{0}/\sigma_{z_{0}})$ & signed transverse (longitudinal) impact parameter significance \\
       $\theta_{\text{rel}} (\phi_{\text{rel}})$ & polar (azimuthal) angle of track with respect to the jet axis \\
       $R$ & angular distance of track and jet axis \\
       $C$ & half-curvature of the track \\
       $m_{\text{ch.}}, q$ & track invariant mass and charge \\
       $\dfrac{|p|_{\text{ch.}}}{|p|_{\text{jet}}}, \ln(|p|_{\text{ch.}}), \ln\left(\dfrac{|p|_{\text{ch.}}}{|p|_{\text{jet}}}\right)$  & (normalised) magnitude of track momentum and logarithms\\
       $\dfrac{E_{\text{ch.}}}{E_{\text{jet}}}, \ln(E_{\text{ch.}}), \ln\left(\dfrac{E_{\text{ch.}}}{E_{\text{jet}}}\right)$  & (normalised) track energy and logarithms\\
       \texttt{isKaon} & if the particle is identified as a $K^{\pm}$ \\
       \texttt{isMuon} & if the particle is identified as a $\mu^{\pm}$ \\
       \texttt{isElectron} & if the particle is identified as an $e^{\pm}$ \\
       \hline
       $\theta_{\text{rel}} (\phi_{\text{rel}})$ & polar (azimuthal) angle of particle with respect to the jet axis \\
       $R$ & angular distance of neutral particle and jet axis \\
       $\dfrac{|p|_{\text{neut.}}}{|p|_{\text{jet}}}, \ln(|p|_{\text{neut.}}), \ln\left(\dfrac{|p|_{\text{neut.}}}{|p|_{\text{jet}}}\right)$  & (normalised) magnitude of particle momentum and logarithms\\
       $\dfrac{E_{\text{neut.}}}{E_{\text{jet}}}, \ln(E_{\text{neut.}}), \ln\left(\dfrac{E_{\text{neut.}}}{E_{\text{jet}}}\right)$  & (normalised) neutral particle energy and logarithms\\
       \texttt{isPhoton} & if the particle is identified as a Photon \\  
              
    \end{tabular}
\caption{\label{tab:jetconstvars}Description of features associated with each jet constituent. The sets of variables are divided into charged particles (tracks) and neutral particles.}
\end{table}

\begin{table}[ht]
\centering
    \begin{tabular}{c|c}
      Input Feature & Description \\
      \hline
      \hline
       
       $|p|,~m$ & 3-momentum magnitude and invariant mass of the SV \\
       $\text{N}_{\text{tracks}}$ & track multiplicity of the SV \\
       $\chi^2,\text{N}_{\text{DoF}}$ & $\chi^2$ and number of degrees of freedom of the SV \\ 
       $\theta_{\text{rel}},\phi_{\text{rel}}$ & polar and azimuthal angle of the SV with respect to the jet axis \\
       $\hat{\text{p}}.\hat{\text{r}}$ & collinearity of SV with respect to PV\\
       $\text{d}_{\text{3D}}, \text{d}_{xy}$ & 3D and transverse distance of the SV from the PV \\
       
    \end{tabular}
\caption{\label{tab:svvars}Description of features associated with each reconstructed secondary vertex. Similar features, with the addition of PDG ID \cite{Workman:2022ynf}, are also defined for V$^0$s while comparing the performance of the tagger trained with and without V$^0$s.}
\end{table}

\begin{figure}%[b]
    \centering
    \begin{subfigure}[t]{0.495\textwidth}
        \centering
        \includegraphics[width=\textwidth]{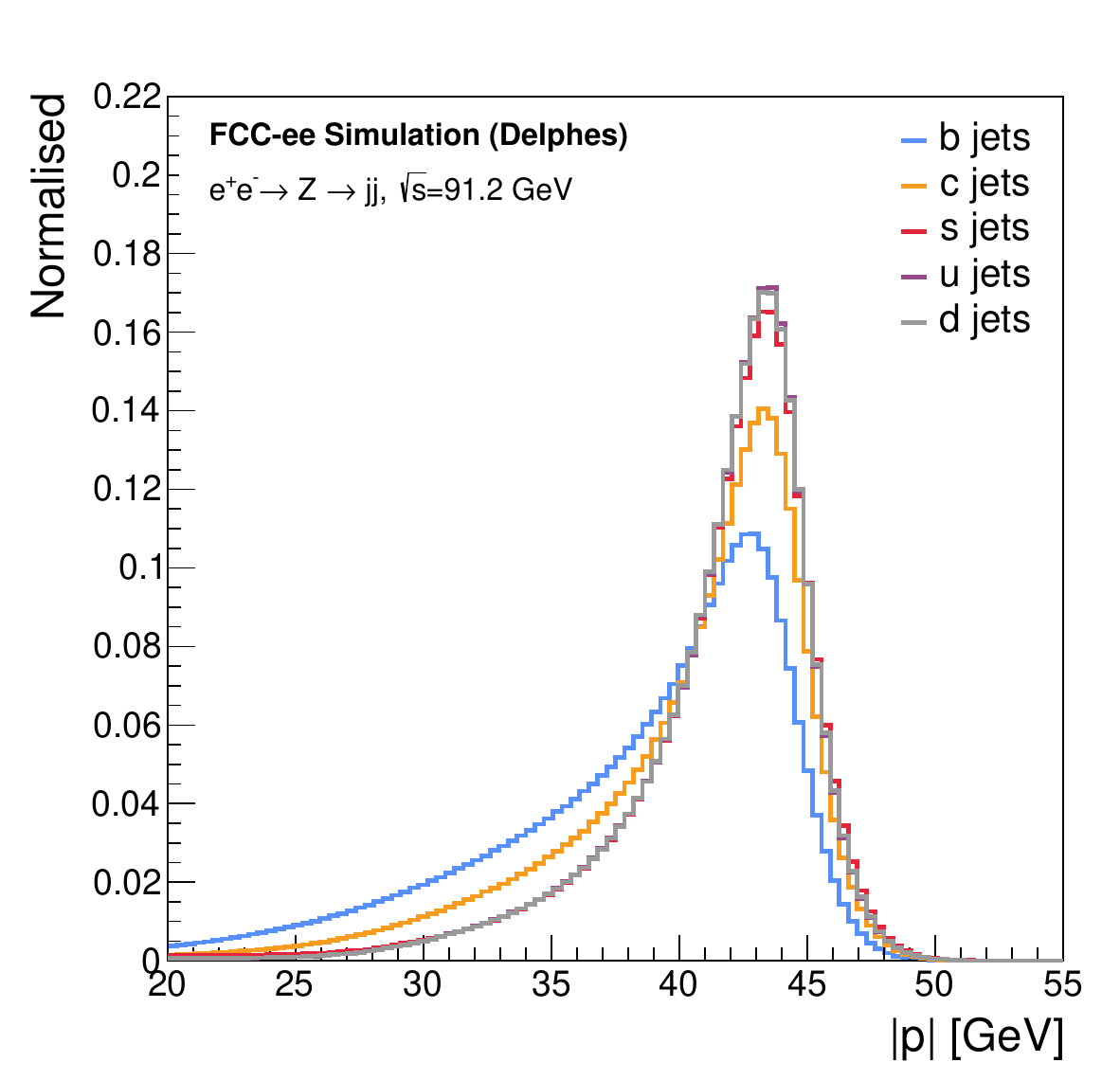}
        \caption{Jet Momentum.}
        \label{fig:jetp}
    \end{subfigure}
    \hfill
    \begin{subfigure}[t]{0.495\textwidth}
        \centering
        \includegraphics[width=\textwidth]{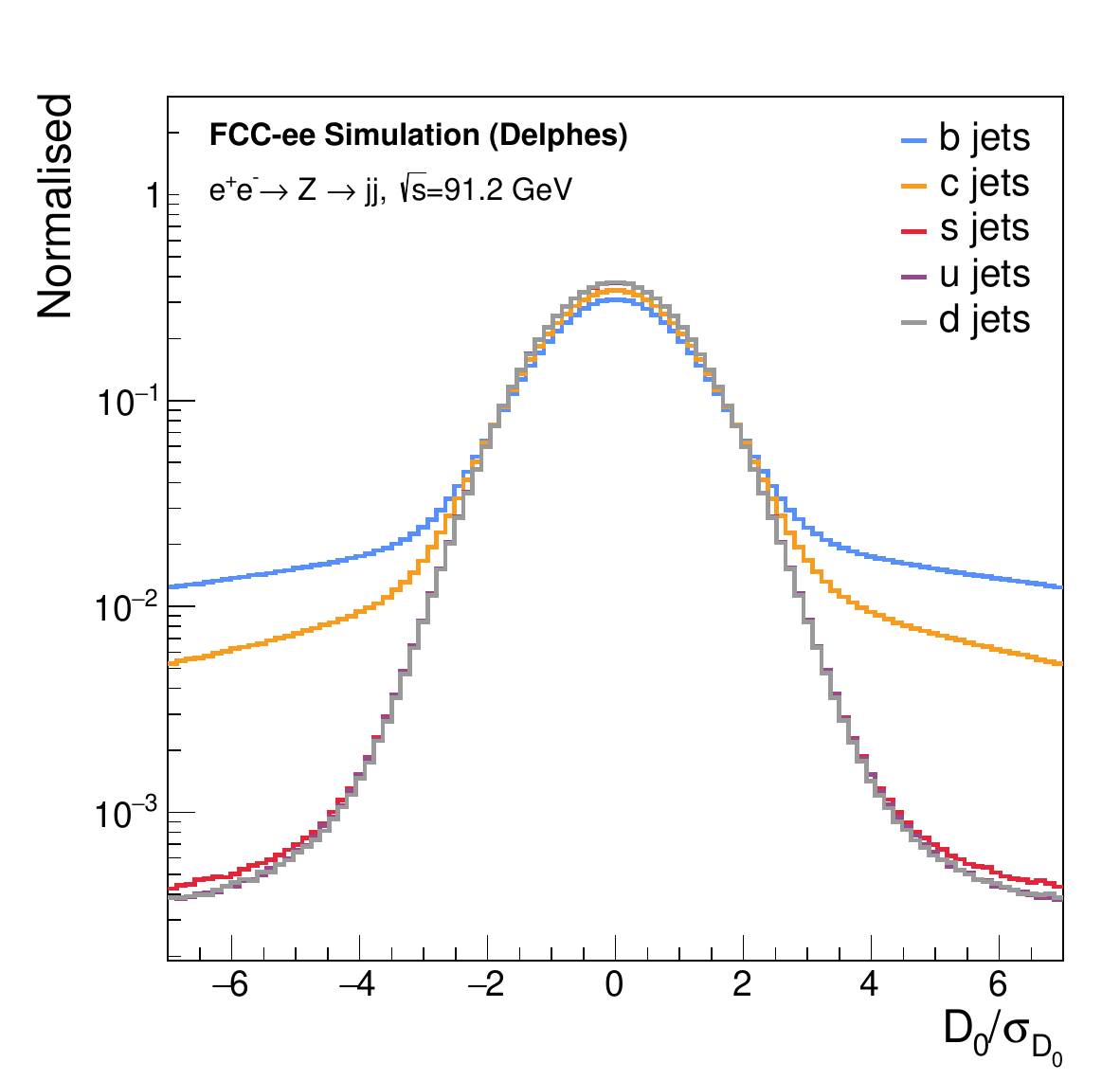}
        \caption{Transverse impact parameter significance.}
        \label{fig:d0sig}
    \end{subfigure}
    \hfill
    \begin{subfigure}[b]{0.495\textwidth}
        \centering
        \includegraphics[width=\textwidth]{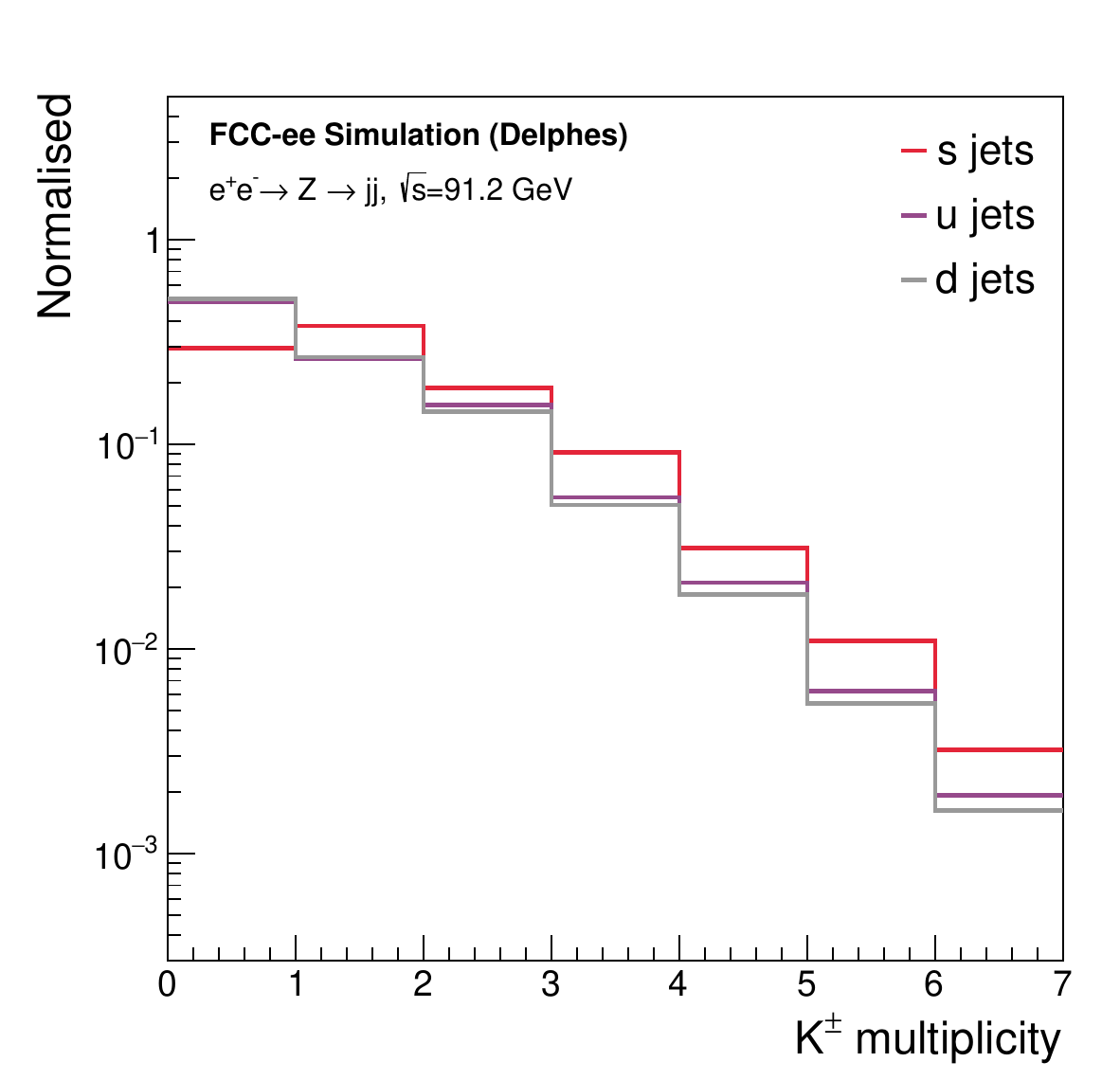}
        \caption{Kaon multiplicities.}
        \label{fig:kaons}
    \end{subfigure}
    \hfill
    \begin{subfigure}[b]{0.495\textwidth}
        \centering
        \includegraphics[width=\textwidth]{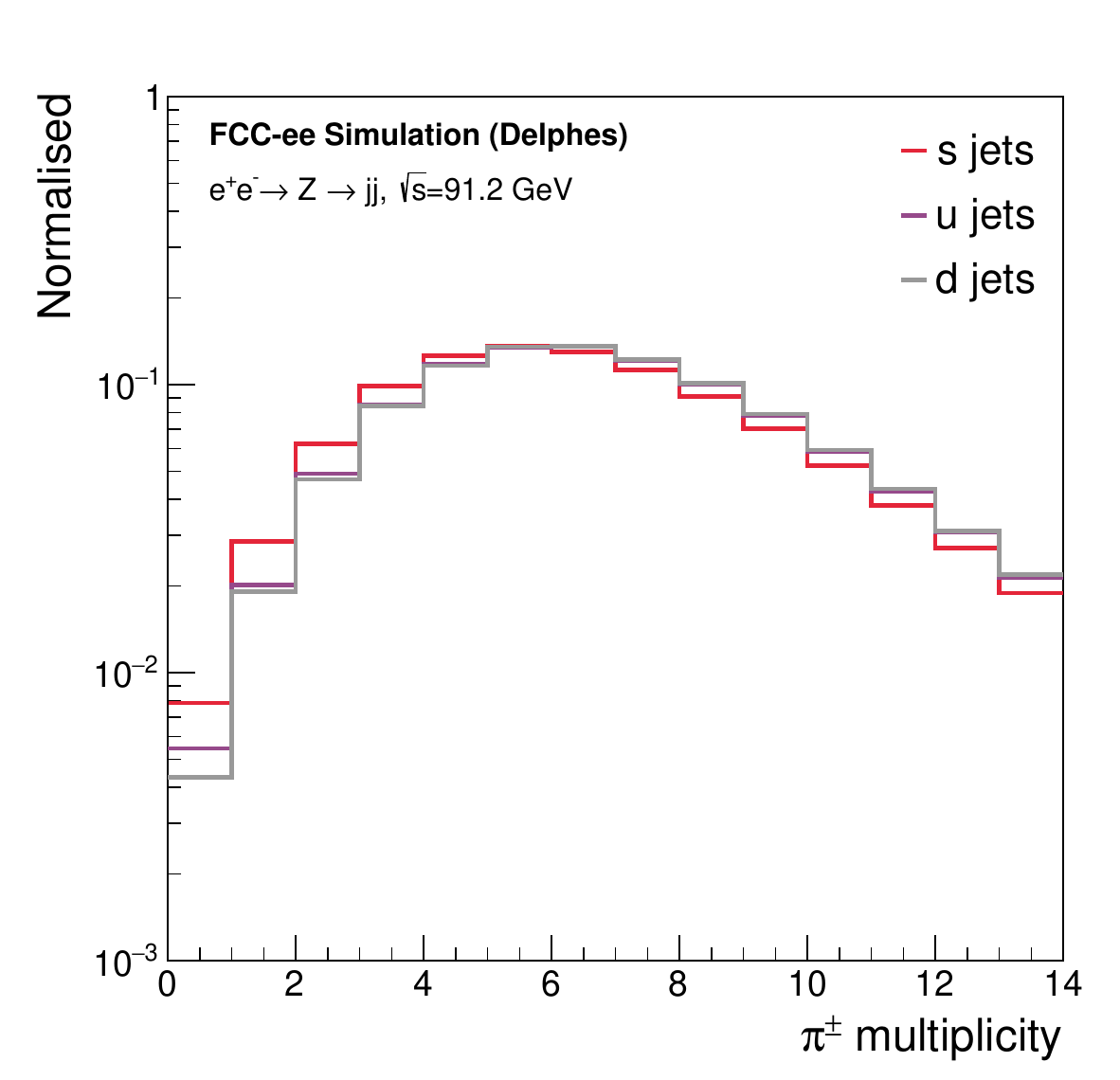}
        \caption{Pion multiplicities.}
        \label{fig:pions}
    \end{subfigure}
    \caption{Distinguishing features in the clustered jets of $e^+e^- \to Z \to q\bar{q}$ events at $\sqrt{s}=91.2$ GeV, separated by flavour. Fig \ref{fig:d0sig} shows a property of the jet constituents, while the rest show properties of the clustered jets. The IDEA detector concept was used for reconstruction.}
    \label{fig:features}
\end{figure}

The jet 3-momentum magnitude distribution of $b$- and $c$-jets tends to be more spread out than that of $s$-, $u$-, and $d$-jets, as seen in Figure \ref{fig:jetp}. This is due to the longer decay chain in $c$-jets than $s$-, $u$-, and $d$-jets, and even longer decay chains in $b$-jets, where more momentum can be lost through neutrinos than in $s$-, $u$-, and $d$-jets.

An important distinguishing variable for $b$-jet identification is the transverse impact parameter ($D_{0}$), which is higher for heavier flavour jets as the decaying $b$ hadrons have a significantly longer lifetime than $c$ or $s$, $u$, $d$ hadrons (except for V$^{0}$s). The differentiating effect between flavours caused by this can be seen more clearly in the transverse impact parameter significance, defined as $S(D_{0}) = D_{0}/\sigma_{D_{0}}$, where $\sigma_{D_{0}}$ is the uncertainty in the measurement of the transverse impact parameter. It is depicted in Figure \ref{fig:d0sig}.

As mentioned in Section \ref{subsec:SV}, $b$-jets tend to have a higher SV multiplicity than $c$-, $s$-, $u$-, and $d$-jets. It is a dominant property in identifying $b$-jets and, to some extent, $c$-jets.

The most challenging background for $s$-tagging is $ud$-jets. Two powerful distinguishing variables tend to be the multiplicities of charged and neutral Kaons and Pions, exploiting the conservation of strangeness during hadronisation in strange jets. These can be seen in Figure \ref{fig:kaons} and \ref{fig:pions}. To distinguish between $K^{\pm}$ and $\pi^{\pm}$, PID techniques like energy loss ($dE/dx$) \cite{Lippmann2012}, ionisation cluster counting ($dN/dx$) \cite{Cataldi:1996mz}, time-of-flight \cite{Klempt:1999tx}, etc. are traditionally used. The $K^{\pm}$/$\pi^{\pm}$ classification is generically emulated, instead of relying on any particular PID technique, with several scenarios of different efficiency to correctly identify $K^{\pm}$, the baseline scenario being $90\%$ efficiency and a $10\%$ efficiency of misidentifying $\pi^{\pm}$ as $K^{\pm}$. The $K^{\pm}$ identification efficiency and the $\pi^{\pm}$ misidentification efficiency are chosen to be constant over the entire momentum range for all the scenarios. The other scenarios considered to study the impact of PID on flavour tagging are summarised in Section \ref{subsec:pid_effect}. The baseline PID scenario was deliberately conservative with respect to the state-of-the-art $K^{\pm}$ identification, which is expected to provide better than $3\sigma$ $K^{\pm}$/$\pi^{\pm}$ separation using cluster counting at FCC-ee \cite{CAPUTO2023167969}, potentially supplemented by time-of-flight \cite{Benedikt:2651299,Wilkinson2021}). This study instead follows PID studies at Belle, which found the average efficiency and fake rate for charged particles between $0.5$ and $4$ GeV/$c$ to be $(87.99 \pm 0.12)\%$ and $(8.53 \pm 0.10)\%$, respectively \cite{NAKANO2002402}. The reconstructed V$^{0}$s, as shown in Figure \ref{fig:V01}, further improve PID by identifying the neutral strange hadrons, $K_{S}^{0}$ and $\Lambda^{0}$. These variables, as described in Table \ref{tab:jetvars}, \ref{tab:jetconstvars}, and \ref{tab:svvars}, are fed into a neural network, the architecture of which is described below.

\subsection{Transformer Models} \label{subsec:Transformer}
Inspired by the success of attention mechanism in Natural Language Processing (NLP) \cite{vaswani2023attentionneed, devlin2019bert} or Computer Vision (CV) \cite{dosovitskiy2021image} tasks, this model adopts Transformer blocks as its primary architectural component. Transformers belong to a class of neural networks that leverage the scaled dot-product attention (SDPA) mechanism \cite{vaswani2023attentionneed}. The attention mechanism enables the model to selectively focus on specific segments of the input sequence while processing each constituent element. In contrast to earlier architectures, such as recurrent models that utilise fixed-size windows or recurrent connections, the attention mechanism dynamically assigns weights to individual elements within the jet based on their relevance, capturing intricate dependencies across the entirety of the jet structure. This adaptive and global weighting scheme empowers the Transformer to effectively model contextual information, a crucial element for understanding and generating coherent high-level features.

\subsubsection{Scaled Dot-Product Attention and Heavy Flavour Transformer Block} \label{subsubsec:SDPA}

The SDPA mechanism uses three inputs: a query matrix $Q$, a key matrix $K$, and a value matrix $V$. In general, the query matrix represents the items for which the attention weights are computed, while the key and value matrices represent all items in the sequence. In this study, the items can be understood to be jet constituents. After being fed into linear layers, the query tensor $Q$ of dimension $(B, N, d_k)$, the key tensor $K$ of dimension $(B, L, d_k)$, and the value tensor $V$ of dimension $(B, L, d_k')$ are fed into the scaled dot-product attention as:
\begin{equation}
\text{Attention}(Q, K, V) = \text{SoftMax}\left( \frac{QK^T}{\sqrt{d_k}}\right) V.
\label{eq:attention}
\end{equation}

The attention mechanism in this study is employed in a specific configuration where the input query, key, and value tensors are identical ($Q=K=V$), and derived from jet constituent features. The tensors Q, K, and V are each passed through linear layers, facilitating the transformation and projection of the input tensors to the attention space. The SDPA is then computed on these transformed tensors as $\text{Attention}(QW^{Q}, KW^{K}, VW^{V})$, where $W^{Q}$, $W^{K}$, $W^{V}$, represent distinct linear transformations. This particular case is commonly referred to as self-attention \cite{vaswani2023attentionneed}.

SDPA is extended to enhance the discriminating power of the model by allowing it to attend to multiple subspaces of attention in parallel. This extension, referred to as Multi-Head Attention (MHA), facilitates the capture of diverse and complementary high-level features from the jet constituent input by projecting the Query, Key, and Value matrices independently for each of the $h$ attention heads. Each attention head performs an SDPA operation, yielding distinct representations. These head representations are then concatenated and passed through a linear layer to integrate the information across heads. The MHA layer can mathematically be represented by the following equations:

\begin{gather}
\text{MHA} (Q, K, V) = \text{Concat}(h_1, ..., h_n)W^O, \\
h_i = \text{Attention}(QW^{Q,i}, KW^{K,i}, VW^{V,i}).    
\label{eq:mha}
\end{gather}

The presented approach, employing the Particle Cloud representation \cite{Qu:2019gqs}, intentionally refrains from employing positional encoding. This decision stems from the absence of a hierarchical structure or positional ordering among the components of the jets, in contrast to sequences such as sentences or image patches. Consequently, the MHA module operates without incorporating positional encoding and instead only leverages permutation invariant mechanisms to capture and process the interrelationships between particles in the jet, yielding meaningful results. The permutation invariance of \texttt{DeepJetTransformer} is established by the properties of permutation equivariance and invariance of function composition \cite{xu2024permutationequivariancetransformersapplications}. The permutation equivariance of each function of the transformer blocks ensures that the network produces a representation of the jet constituents respecting the Particle Cloud properties. It is made sure that the network's flavour predictions remain invariant under the permutation of jet constituents by applying a permutation invariant attention pooling followed by linear layers for classification. By analogy with graph structures, the attention mechanism can be interpreted similarly to the ones used in fully connected graph networks, with the attention scores playing a role similar to the edge features by capturing relationships within the jet structure.

After establishing the fundamental components of the utilised model's architecture, the foundational block forming the backbone of the model can be defined. This essential building block, referred to as the Heavy Flavour Transformer block (HFT), is structured in the following manner:

\begin{itemize}
\item The jet constituent inputs are fed into a basic Multilayer Perceptron (MLP) layer followed by a ReLU activation function.
\item The product of the MLP layer is then fed in an MHA layer before using a residual connection and layer normalisation.
\item In addition to the MHA layer, a fully connected feed-forward layer is also added, identical to the original Transformer implementation \cite{vaswani2023attentionneed} followed by a last residual connection and layer normalisation.
\end{itemize}

Unlike other Transformer models applied to jet (sub)structures \cite{Mondal_2024, Qu:2022mxj}, a $cls$ token is not employed to embed the information of the jet structures into relevant features for classification. Instead, an attention pooling is introduced, behaving similarly to a Max or Average pooling layer with an attention mechanism and learnable parameters. The attention pooling operates by employing an MLP projection layer, which enables local feature extraction. Subsequently, a softmax activation function is applied to calculate attention weights, allowing the layer to emphasise relevant elements in the sequence. The attention weights are then used to aggregate the sequence information by performing a weighted sum. To enhance the layer's performance, batch normalisation is applied, the ReLU~\cite{4082265} activation function is used to introduce non-linearity, and dropout regularisation is incorporated to prevent overfitting. The attention pooling layer can effectively capture essential information from the sequence and produce a condensed representation by incorporating these components that can be utilised for jet flavour classification. In the context of jet flavour tagging, Transformer models can be interpreted as fully connected graph networks using the jet’s constituents as the nodes, and the SDPA as a mechanism connecting all the node information for enhancing the feature engineering of the model.

\begin{figure}
    \centering
    \includegraphics[width=1.\textwidth]{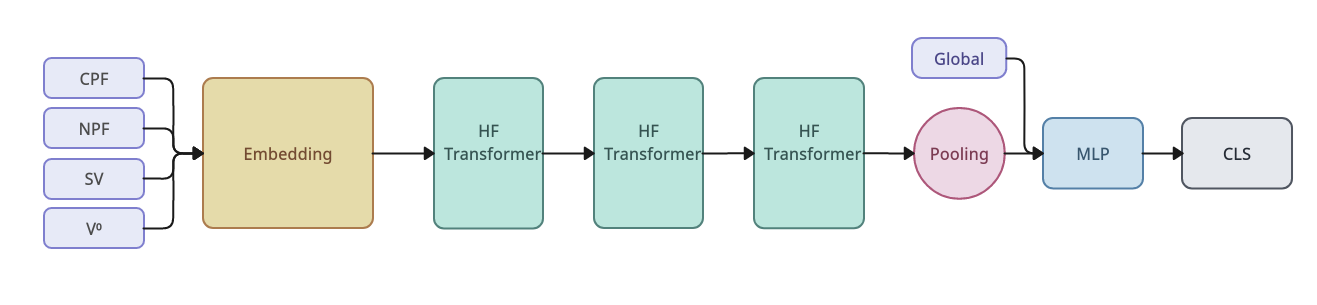}
    \caption{Schematic structure of \texttt{DeepJetTranformer} model.}
    \label{fig:DJT_schematic}
\end{figure}

\subsubsection{DeepJetTransformer Architecture} \label{subsubsec:DJT}

With all the components of \texttt{DeepJetTransformer} defined, the global structure of the model can be described. Figure \ref{fig:DJT_schematic} illustrates the detailed structure of \texttt{DeepJetTransformer}, which is as follows:

\begin{itemize}
\item[-] The features of distinct jet constituents first undergo embedding via a series of three MLPs with output feature dimensions of (64, 128, 128), employing ReLU activation, residual connections, and batch normalisation. Dropout regularisation with a rate of 0.1 is applied following each batch normalisation operation.
\item[-] The resulting feature tensors are then concatenated to form a single tensor containing all the comprehensive information of the jet constituents.
\item[-] This global tensor is subsequently passed through three HFT blocks, each possessing a feature dimension of 128. Each block contains eight attention heads and incorporates a dropout rate of 0.1.
\item[-] The representation of the jet structure, obtained through the HFT blocks, is further condensed via attention pooling. The resulting tensor is concatenated with jet-level features, yielding a vector containing 135 relevant features for heavy flavour classification. Among these, 128 features originate from attention pooling, while the remaining seven variables represent the jet-level attributes.
\item[-] The jet representation is subsequently fed to three MLPs with output feature dimensions of (135, 135, 135), mirroring the structure of the input embedding MLPs.
\item[-] A single MLP followed by a SoftMax function is applied finally for classification.
\end{itemize}

In summary, the three main differences in the architectures of \texttt{DeepJetTransformer} and \texttt{ParT (plain)} \cite{Qu:2022mxj}, which both implement a pure Transformer architecture, are the use of attention pooling instead of the typical $cls$ token, additional linear layers prior to the MHA, and the inclusion of the of jet-level variables (listed in Table \ref{tab:jetvars}) in addition to jet constituent variables.

\subsubsection{Training Methodology} \label{subsubsec:training}

\texttt{PyTorch (v1.10.1)} \cite{NEURIPS2019_9015} was employed as the deep learning library in this study for the neural network model construction and the training process. The optimiser utilised was the Lookahead optimiser \cite{zhang2019lookahead}, with hyperparameters $k=6$ and $\alpha$ = 0.5 and a RAdam \cite{liu2021variance} as the base optimiser with a learning rate of 5e-3 and decay rates ($\beta_1,~\beta_2$) set to $(0.95,~0.999)$. The training was conducted over 70 epochs with a batch size of 4000, accompanied by a per-epoch linear learning rate decay starting after 70\% of the training, gradually decreasing to 5e-5 by the final epoch. A cross-entropy loss function was used for optimisation. The training dataset comprised of 1 million jets, divided into an $80/20\%$ train-validation split. Finally, the model was evaluated on a separate dataset of 1 million jets for performance assessment. Documentation for the sample preparation and training methodology, along with the relevant code, is publicly available here: \href{https://github.com/Edler1/DeepJetFCC/tree/master/docs}{DeepJetFCC}\footnote{\href{https://github.com/Edler1/DeepJetFCC/tree/master/docs}{https://github.com/Edler1/DeepJetFCC/tree/master/docs}}.

\section{Classifier Performance} \label{sec:performance}
To evaluate the performance of \texttt{DeepJetTransformer}, clustered jets from $Z\rightarrow q \bar{q}$ events at $\sqrt{s}=91.2$ GeV and $Z(\to\nu \nu)H(\to q \bar{q})$ events at $\sqrt{s}=240$ GeV were considered. The tagger was trained separately for each process. The emphasis was placed on the $Z$ resonance for these studies, with the classification of $H \rightarrow q \bar{q}$ events serving primarily as a comparison to the classification performance of other jet flavour taggers for future colliders, like \texttt{ParticleNetIDEA} \cite{Bedeschi_2022, Gautam:2022szi}.
A binary classifier was constructed for each jet flavour $q \equiv u,d,s,c,b,(g)$ with a signal flavour ($i$) and a background flavour ($j$):
\begin{equation}
S_{ij} = \frac{S_{i}}{S_{i} + S_{j}},
\label{eq:discriminant}
\end{equation} 
where $S_{i}$ are the outputs of the classifier normalised using the SoftMax function, as described in Section \ref{subsubsec:DJT}. These normalised outputs, which are constrained to lie between 0 and 1 and sum to 1 across all jet flavours, are referred to as "softmaxed classifier outputs" for brevity.

\begin{figure}[t]
    \centering
    \begin{subfigure}[t]{0.495\textwidth}
        \centering
        \includegraphics[width=\textwidth]{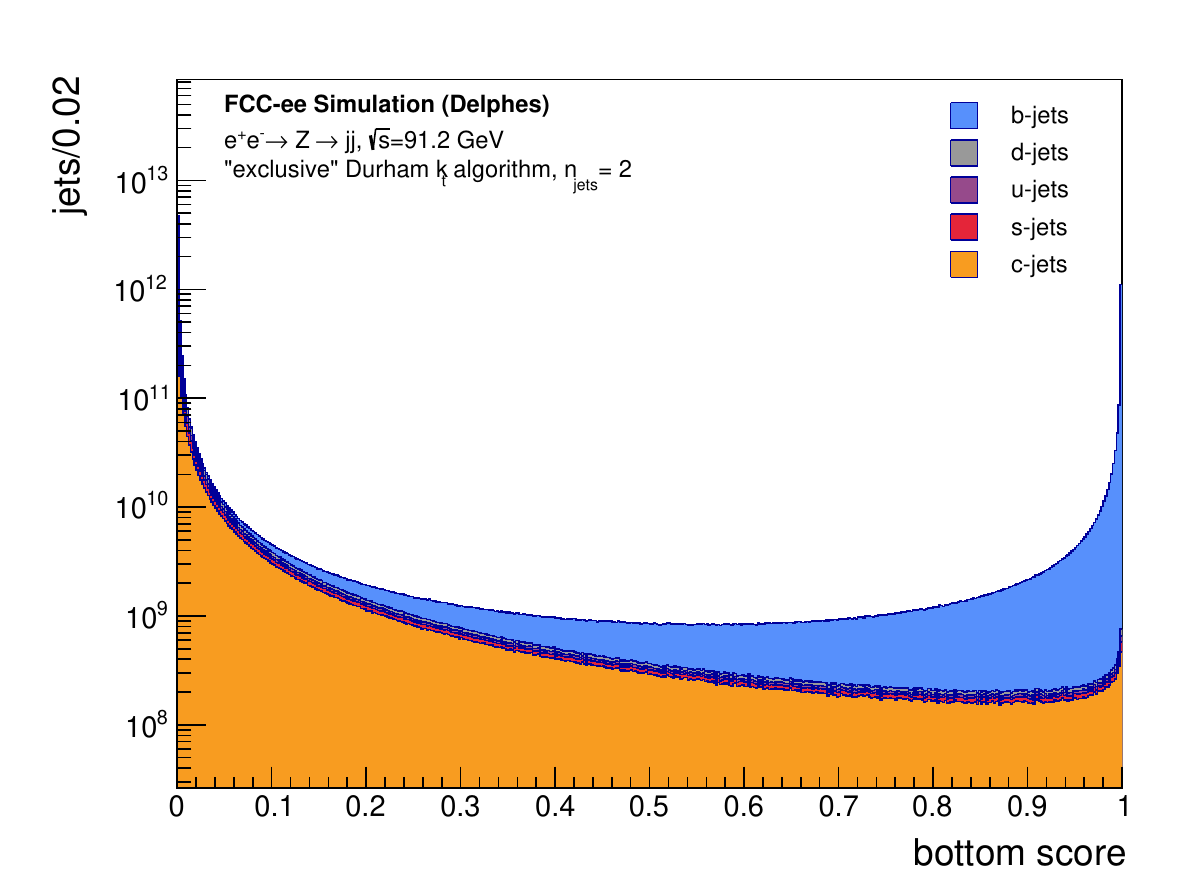}
        \caption{}
        \label{fig:DJT_b_disc}
    \end{subfigure}
    \hfill
    \begin{subfigure}[t]{0.495\textwidth}
        \centering
        \includegraphics[width=\textwidth]{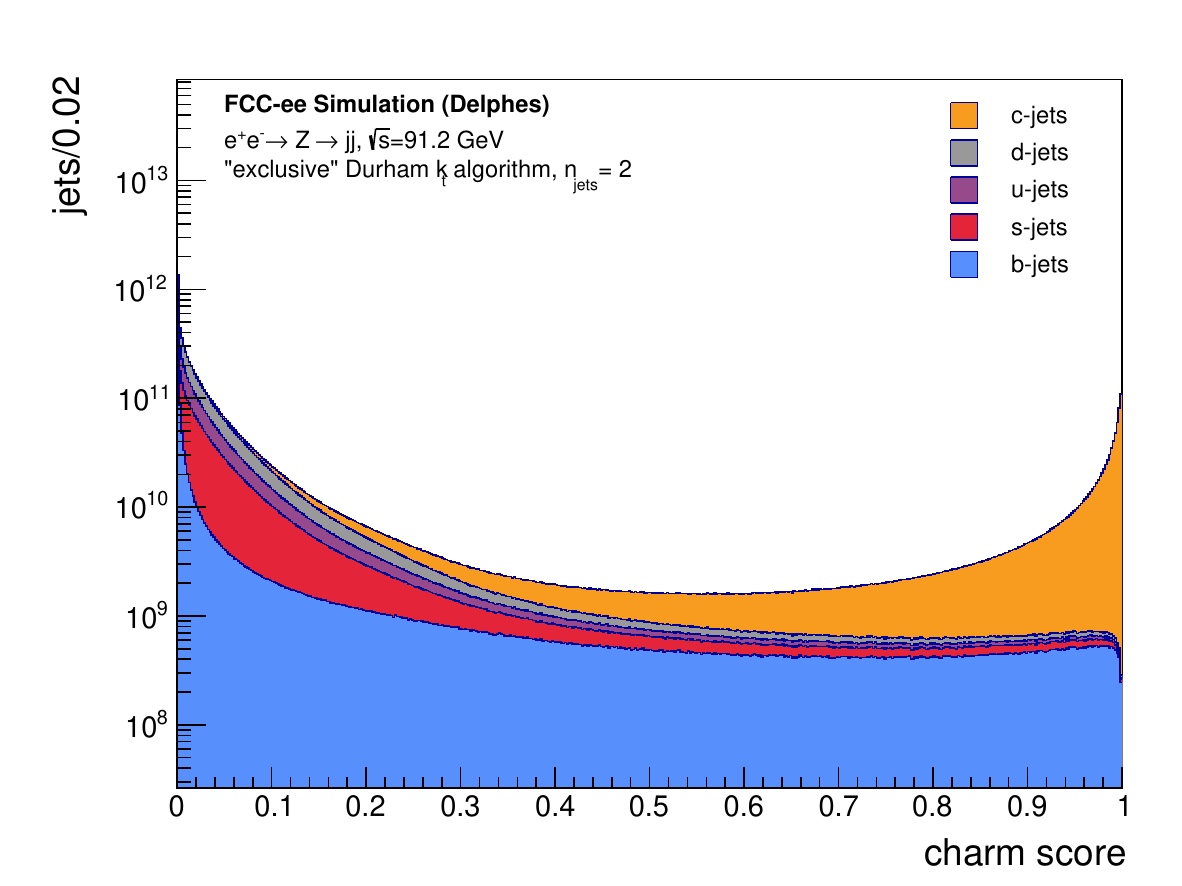}
        \caption{}
        \label{fig:DJT_c_disc}
    \end{subfigure}
    \hfill
    \begin{subfigure}[t]{0.495\textwidth}
        \centering
        \includegraphics[width=\textwidth]{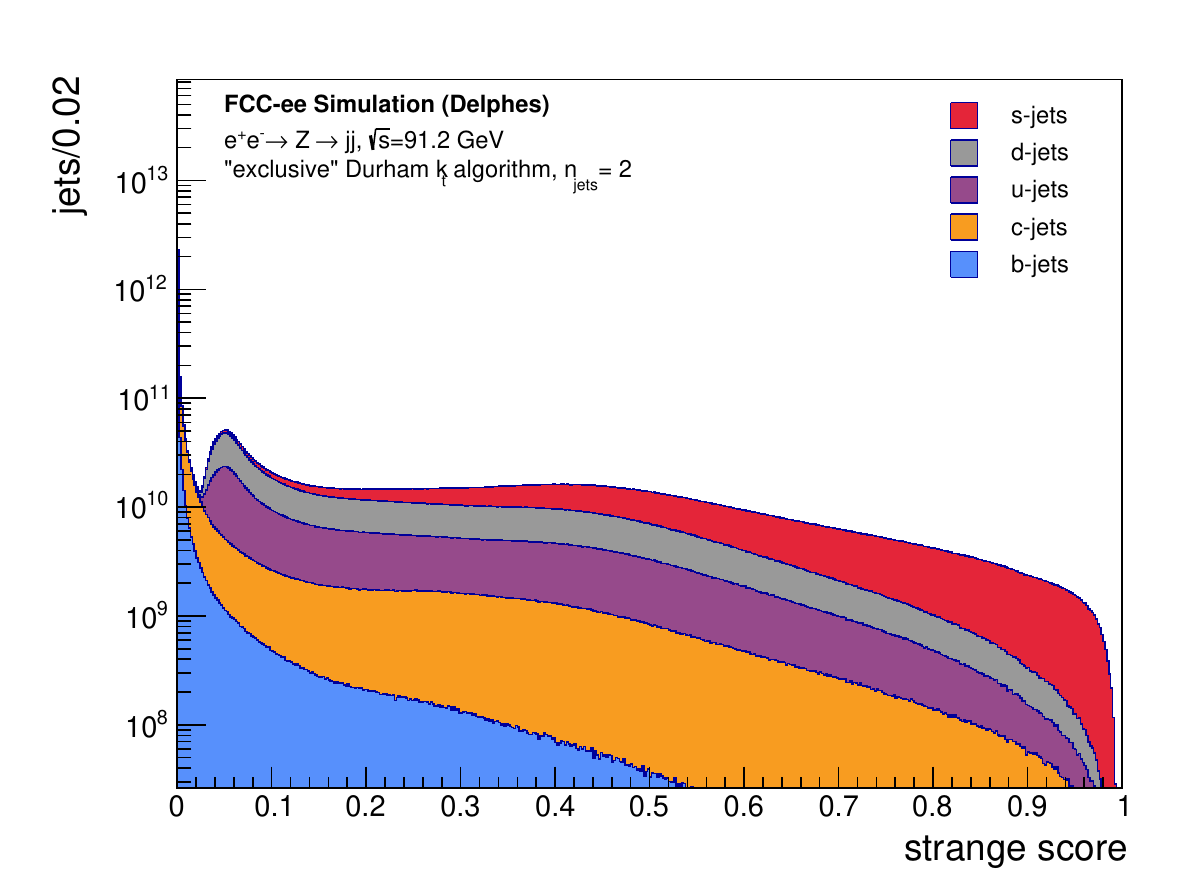}
        \caption{}
        \label{fig:DJT_s_disc}
    \end{subfigure}
    \hfill
    \begin{subfigure}[t]{0.495\textwidth}
        \centering
        \includegraphics[width=\textwidth]{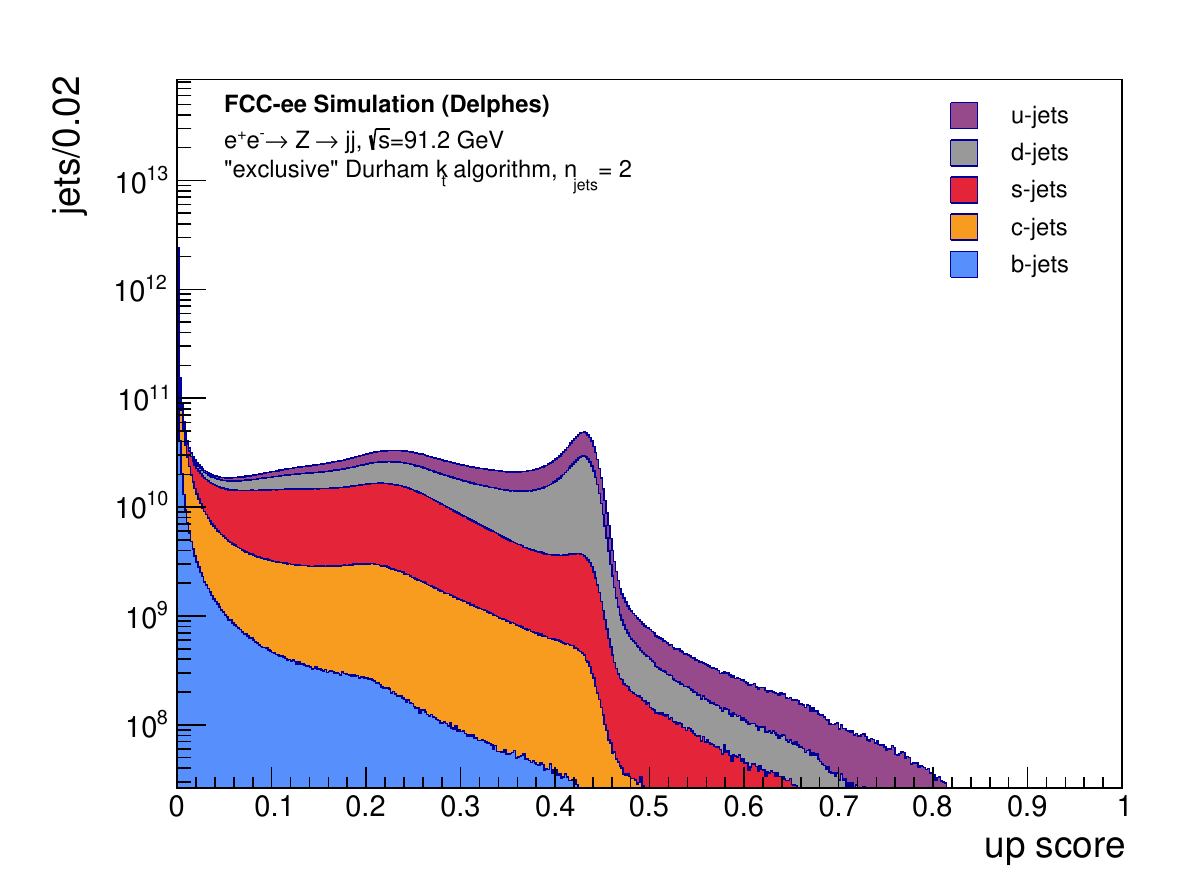}
        \caption{}
        \label{fig:DJT_u_disc}
    \end{subfigure}
    \hfill
    \begin{subfigure}[t]{0.495\textwidth}
        \centering
        \includegraphics[width=\textwidth]{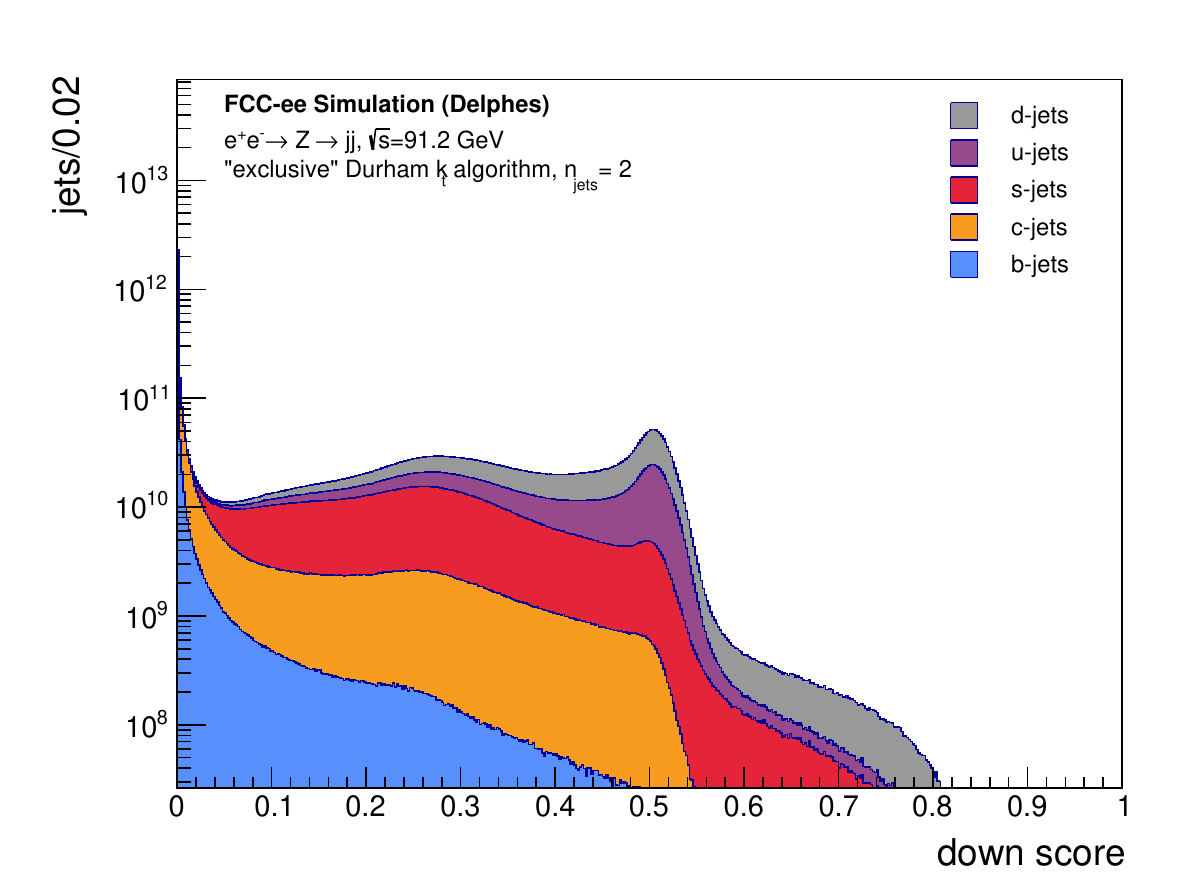}
        \caption{}
        \label{fig:DJT_d_disc}
    \end{subfigure}
    \caption{\label{fig:DJT_discriminants}The softmaxed classifier outputs ($S_i$) of the five output nodes of \texttt{DeepJetTransformer} trained with clustered jets of $e^+e^- \to Z \to q\bar{q}$ events at $\sqrt{s}=91.2$ GeV. The contributions of different MC flavours have been displayed.}
\end{figure}

\begin{figure}%[t]
    \centering
    \begin{subfigure}[t]{0.495\textwidth}
        \centering
        \includegraphics[width=\textwidth]{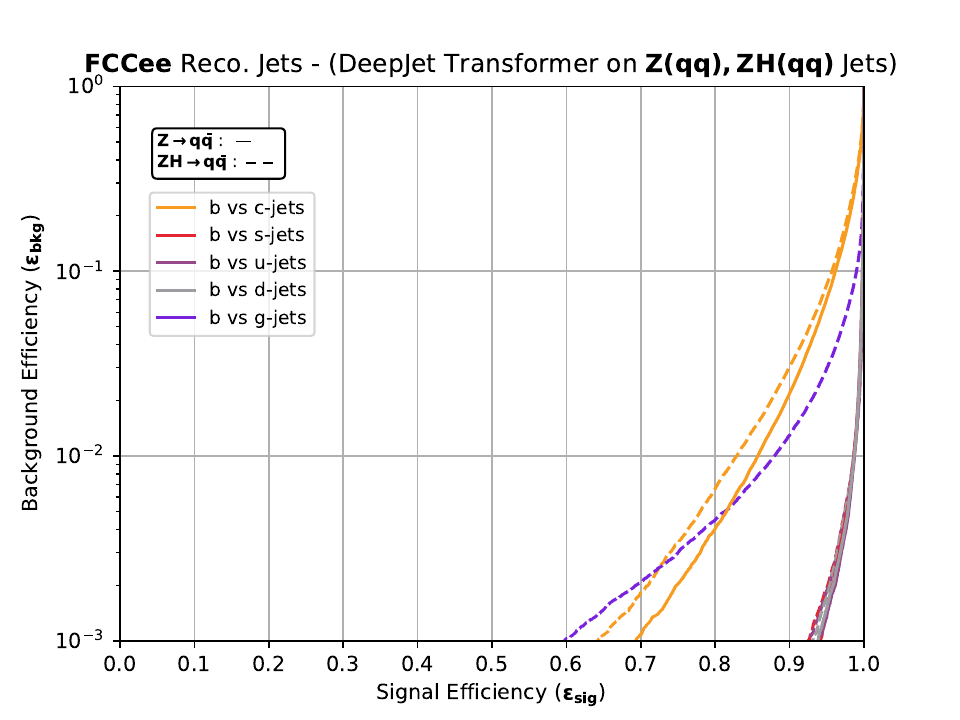}
        \caption{bottom tagging}
        \label{fig:ROC_b_log}
    \end{subfigure}
    \hfill
    \begin{subfigure}[t]{0.495\textwidth}
        \centering
        \includegraphics[width=\textwidth]{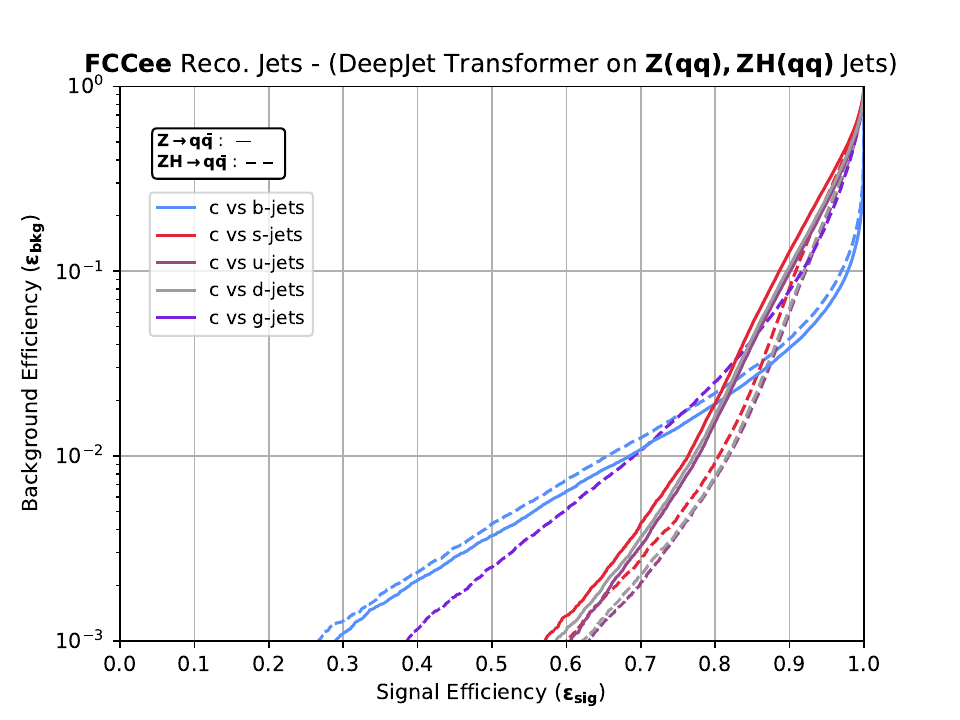}
        \caption{charm tagging}
        \label{fig:ROC_c_log}
    \end{subfigure}
    \hfill
    \begin{subfigure}[t]{0.495\textwidth}
        \centering
        \includegraphics[width=\textwidth]{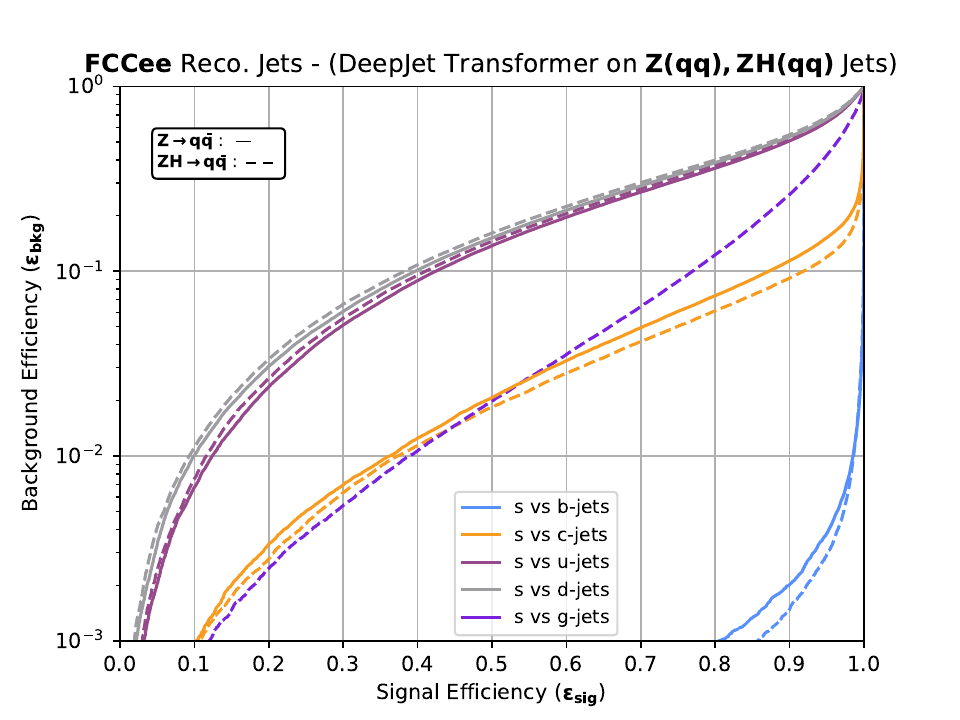}
        \caption{strange tagging}
        \label{fig:ROC_s_log}
    \end{subfigure}
    \hfill
    \begin{subfigure}[t]{0.495\textwidth}
        \centering
        \includegraphics[width=\textwidth]{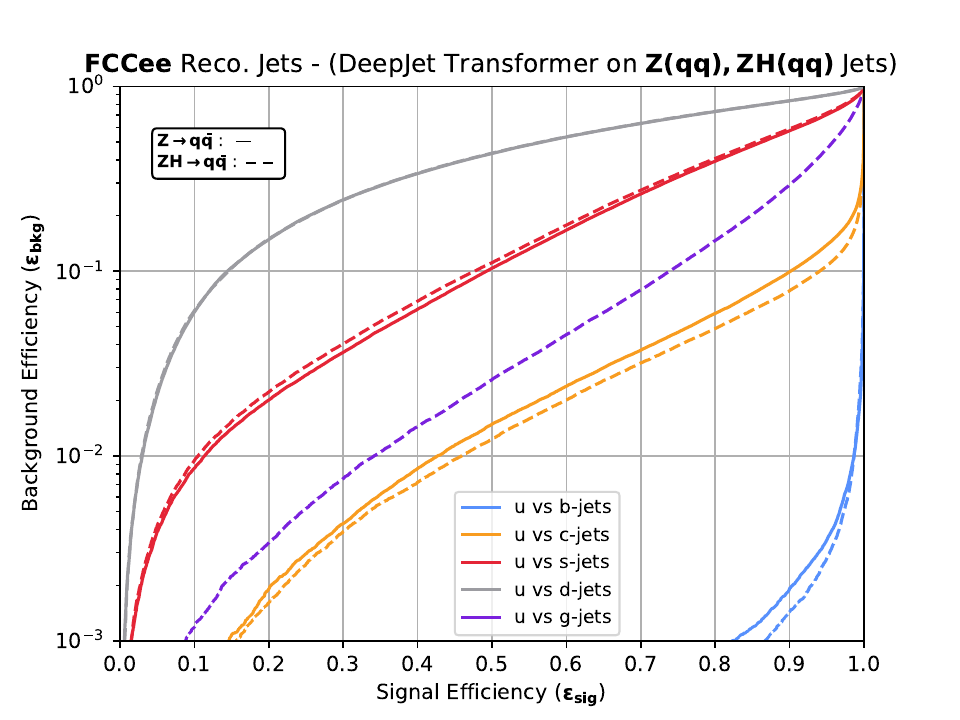}
        \caption{up tagging}
        \label{fig:ROC_u_log}
    \end{subfigure}
    \hfill
    \begin{subfigure}[t]{0.495\textwidth}
        \centering
        \includegraphics[width=\textwidth]{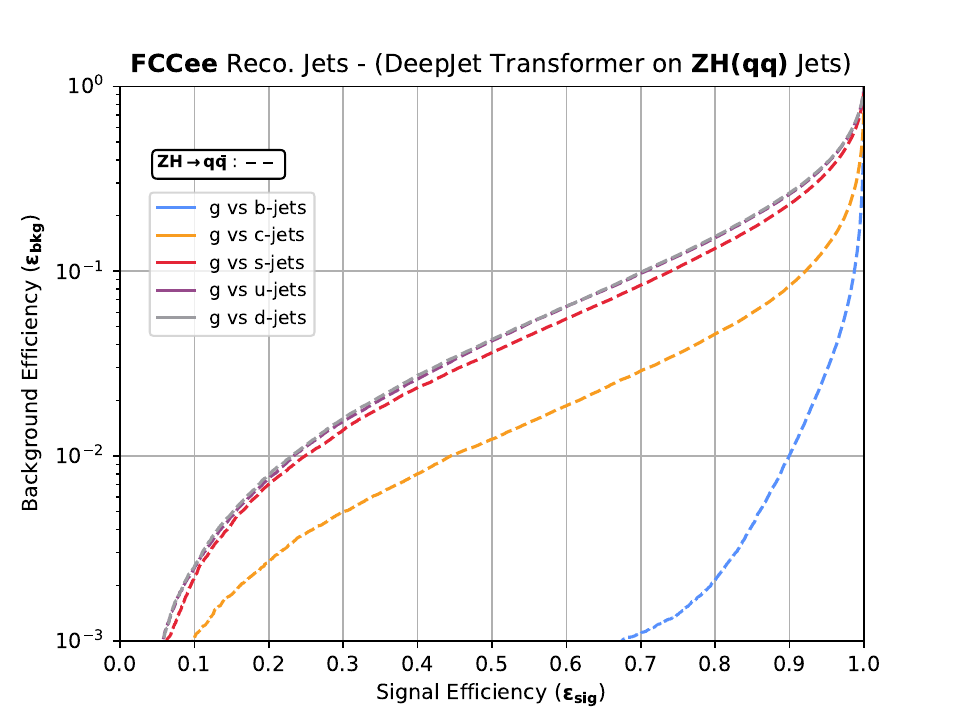}
        \caption{gluon tagging}
        \label{fig:ROC_g_log}
    \end{subfigure}
    \caption{\label{fig:ROC_all_flav}ROC curves for each $S_{ij}$ combination, as defined in Eq. \ref{eq:discriminant}, where $i$ is the signal parton flavour and $j$ is the background flavour. The solid lines correspond to the classification of jets at the $Z$ resonance at $\sqrt{s}=91.2$ GeV, while the dashed lines correspond to the classification of jets from $Z(\to\nu\nu)H(\to q\bar{q})$ events at $\sqrt{s}=240$ GeV. The tagger was trained separately for each process. No quark-gluon discrimination results are presented for jets from $Z\to q\bar{q}$ events as the $Z$ boson does not decay into gluons.}
\end{figure}

The five softmaxed classifier outputs of \texttt{DeepJetTransformer} are shown in Figure \ref{fig:DJT_discriminants}. ROC curves were computed for each $S_{ij}$ combination and are depicted in Figure \ref{fig:ROC_all_flav} for the $Z$ resonance and the $ZH$ training. Predictably, the strongest discrimination is between $b$-jets and $s$-, $u$-, $d$- jets and is roughly equivalent for all three background jets. The dominant background is from $c$-jets, originating from the similarity of $b$- and $c$-jets with a single reconstructed SV. Discriminating $c$-jets from $u$-, $d$- and $s$-jets exhibits similar performances, with relatively worse discrimination of the $s$-jet background. Figure \ref{fig:ROC_c_log} shows that as the efficiency increases from the right to the left side of the plot, $s$-, $u$- and $d$-jets are discriminated worse than $b$-jets in the high-efficiency regime for $c$-jets until a turnover point at $\epsilon_{sig}^{c} \approx 80\%$, after which distinguishing $s$-, $u$- and $d$-jets becomes considerably easier than $b$-jets. Such a turnover can also be found in \texttt{ParticleNetIDEA} \cite{Bedeschi_2022}. The sub-leading background comes from $s$-jets, clustered at low to mid charm scores, as also evident in Figure \ref{fig:DJT_c_disc}, primarily as no SVs can be reconstructed for a significant number of $c$-jets, leaving few variables to distinguish $c$- and $s$-jets.

When $s$-jets are taken to be the signal, as shown in Figure \ref{fig:ROC_s_log}, $c$- and $ud$-jets present the most challenging backgrounds, with $c$-jets being easier to discriminate against at all signal purities. The $c$-jet background comes from jets where a charm hadron decays to a strange hadron, and only the V$^{0}$ can be reconstructed, or a strange hadron carries excess momentum. Some discrimination against the dominant $ud$-jets background can be achieved at higher cuts on the strange score, owing to the $K^{\pm}/\pi^{\pm}$ separation and V$^{0}$ reconstruction. Finally, Figures \ref{fig:DJT_u_disc} and \ref{fig:DJT_d_disc} show almost overlapping distributions of classifier scores for $u$- and $d$-jets. Figure \ref{fig:ROC_u_log} validates that classification is most challenging for $u$- and $d$-jets. When $u$-jets are taken to be the signal, it can be seen that \texttt{DeepJetTransformer} learns to discriminate $u$- vs $d$-jets with a $\epsilon_{sig}^{u} \approx 15\%$ at a $\epsilon_{bkg} = 10\%$, which is better than a random classifier, although not considerably. The discrimination is likely related to a mapping to the initiating parton’s charge, such as the jet charge \cite{Field:1977fa, Krohn:2012fg}, the effect of which is diluted by the presence of antiquarks.

While considering the performance for $H(\to q \bar{q})$ jets, depicted as dashed lines in Figure \ref{fig:ROC_all_flav}, no clear trend can be observed. Slight degradation in performance can be observed in the case of $b$ tagging, compared to $Z\rightarrow q \bar{q}$ jets, particularly when $c$-jets are taken to be the background. The discrimination of $c$-jets vs $s$-, $u$-, and $d$-jets is found to perform relatively the best with respect to the $Z\rightarrow q \bar{q}$ jets when considering the percent-improvement in the ROC Area Under the Curve metric.

Figure \ref{fig:ROC_g_log} shows that the best quark-gluon discrimination can be achieved against the $b$ quarks. This performance can be attributed to several discriminating variables, like jet-constituent multiplicity, constituent momentum distribution, etc., but is dominated by the presence or absence of reconstructed SVs. It is the most challenging to discriminate the $s$, $u$, and $d$ quarks from gluons due to their similar jet composition. 

\begin{figure}[ht]
    \centering
    \begin{subfigure}[t]{0.495\textwidth}
        \centering
        \includegraphics[width=\textwidth]{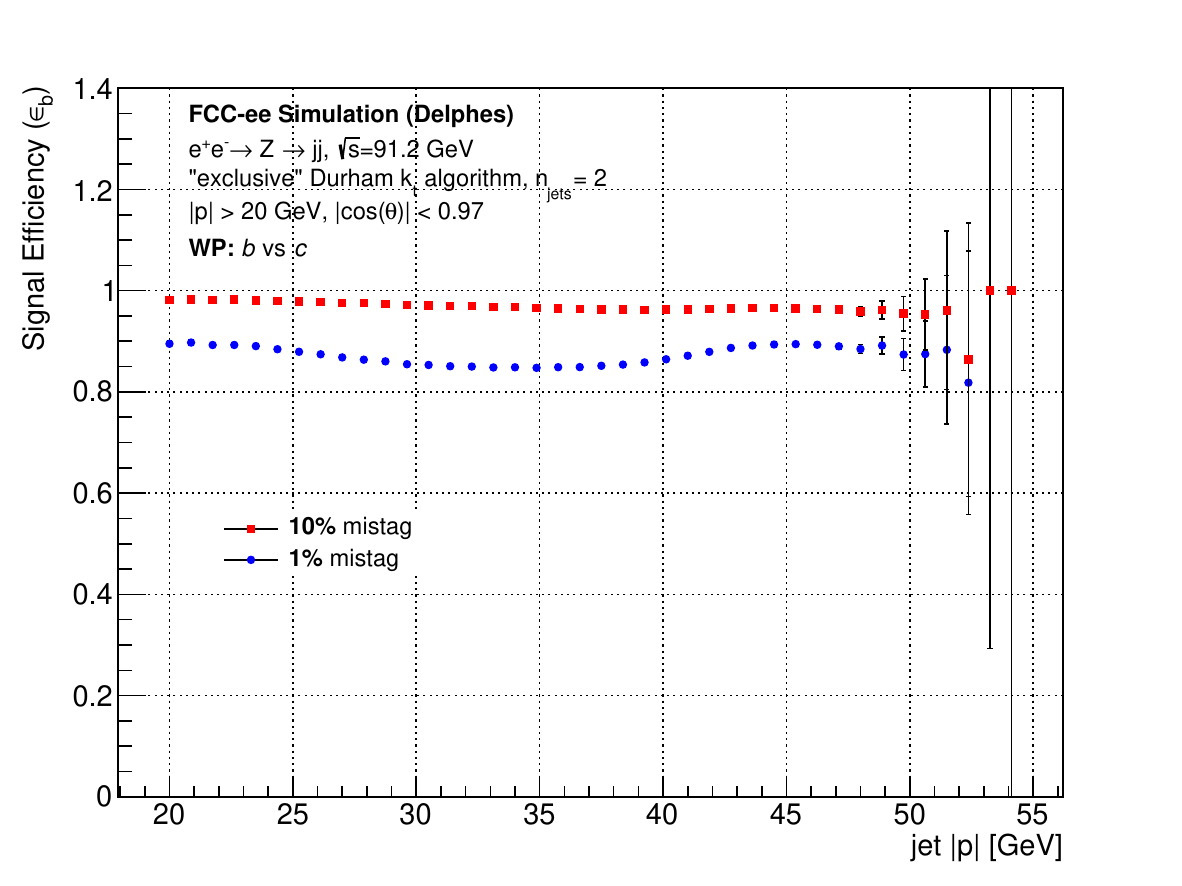}
        \caption{\textit{b} vs \textit{c}}
        \label{fig:DJT_bVc_eff_p}
    \end{subfigure}
    \hfill
    \begin{subfigure}[t]{0.495\textwidth}
        \centering
        \includegraphics[width=\textwidth]{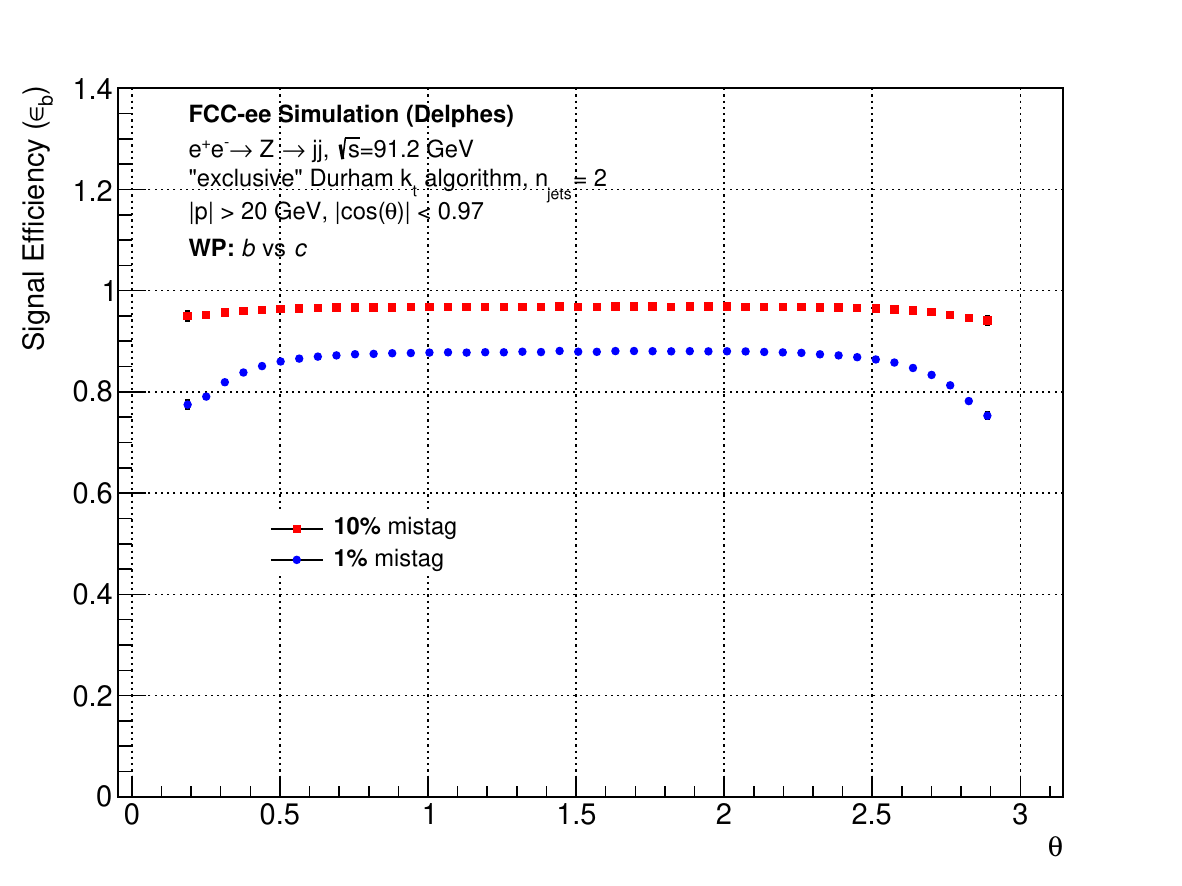}
        \caption{\textit{b} vs \textit{c}}
        \label{fig:DJT_bVc_eff_theta}
    \end{subfigure}
    \hfill
    \begin{subfigure}[t]{0.495\textwidth}
        \centering
        \includegraphics[width=\textwidth]{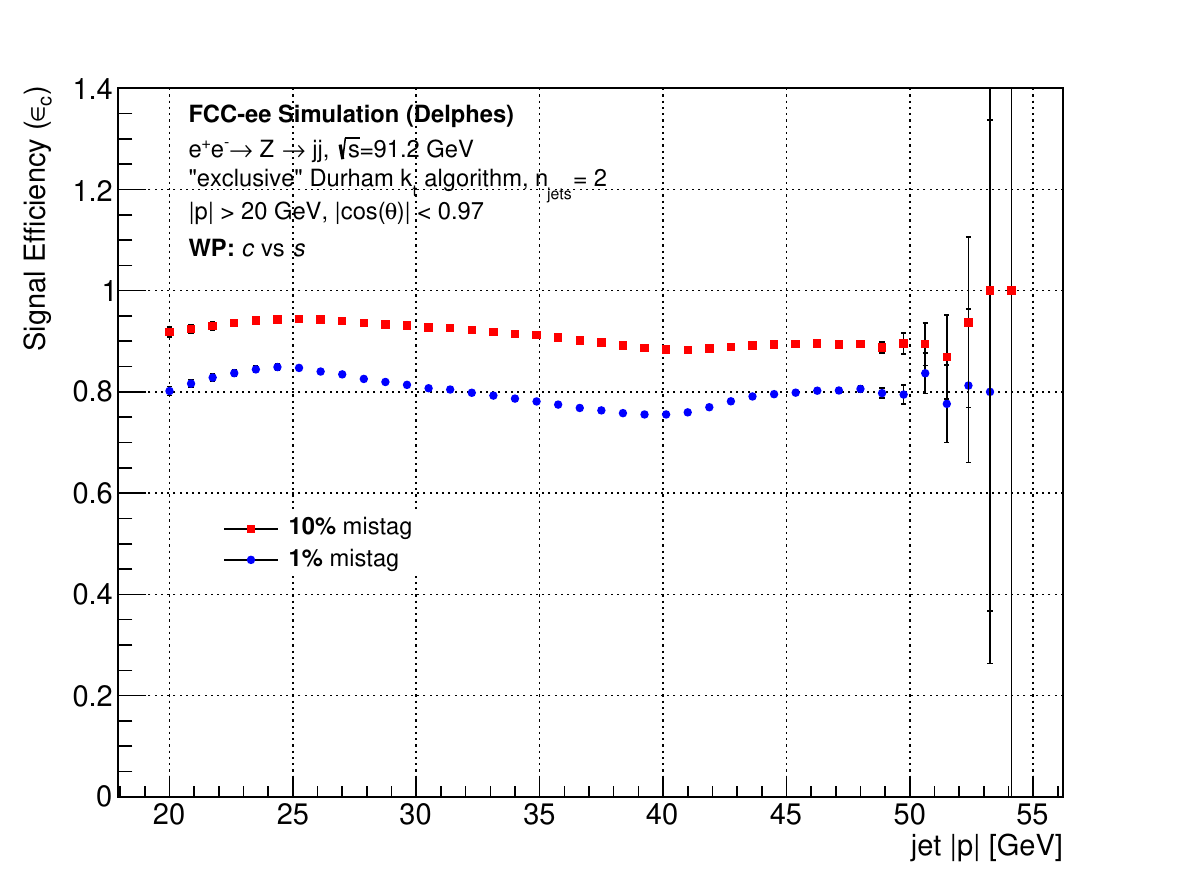}
        \caption{\textit{c} vs \textit{s}}
        \label{fig:DJT_cVs_eff_p}
    \end{subfigure}
    \hfill
    \begin{subfigure}[t]{0.495\textwidth}
        \centering
        \includegraphics[width=\textwidth]{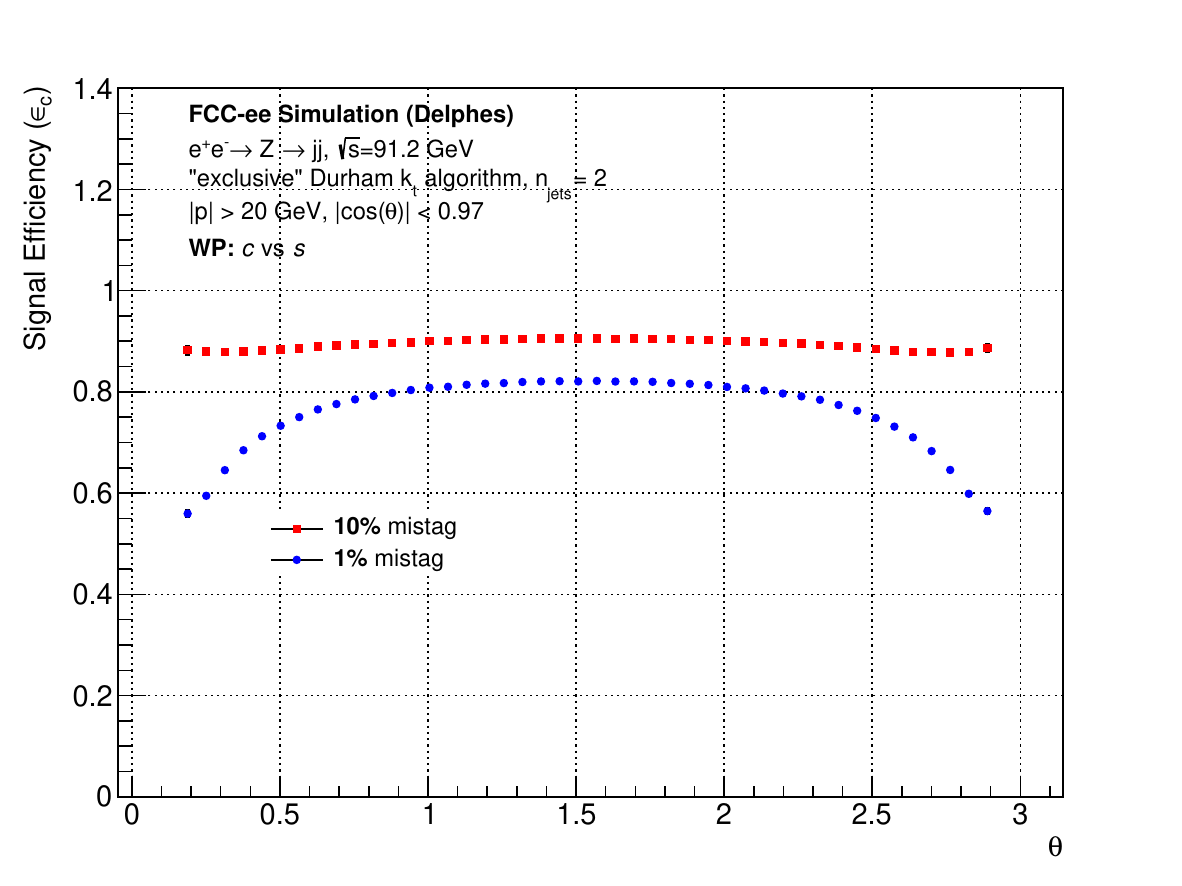}
        \caption{\textit{c} vs \textit{s}}
        \label{fig:DJT_cVs_eff_theta}
    \end{subfigure}
    \hfill
    \begin{subfigure}[t]{0.495\textwidth}
        \centering
        \includegraphics[width=\textwidth]{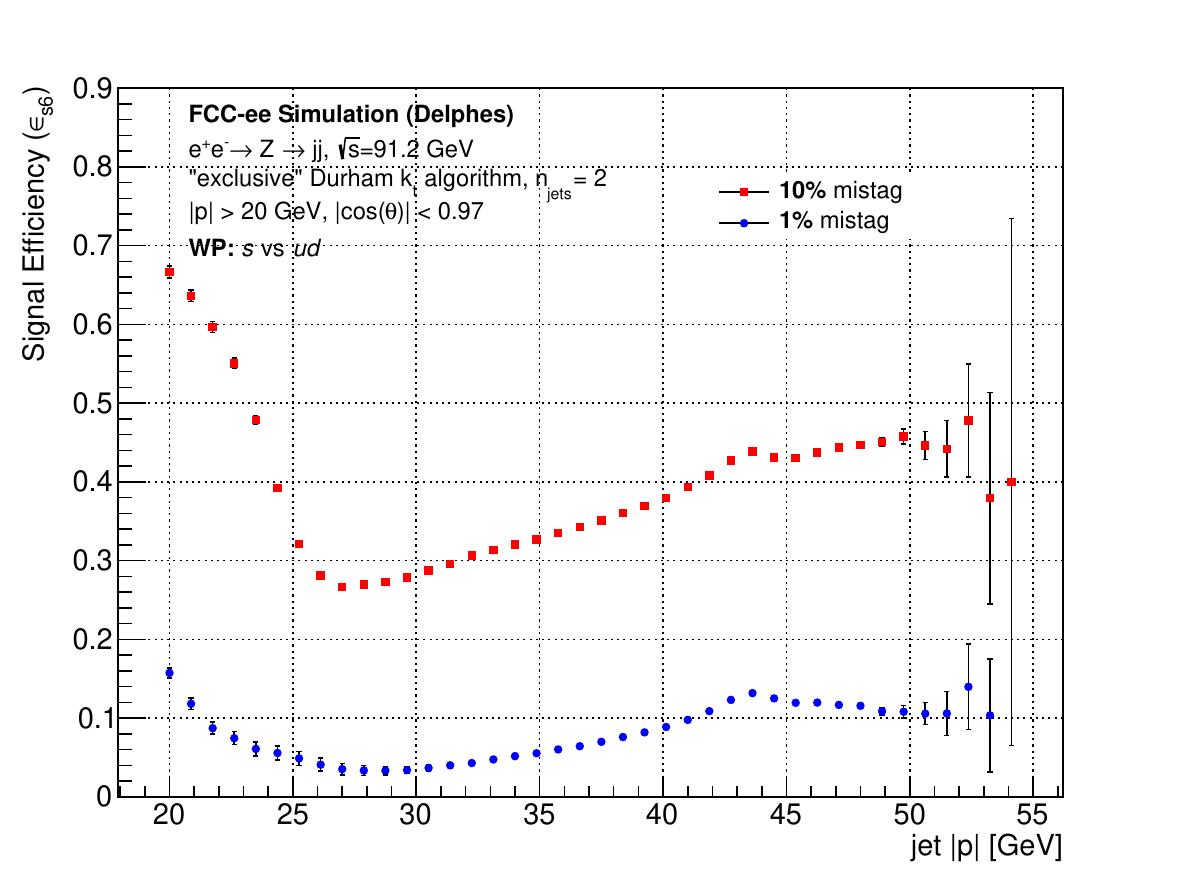}
        \caption{\textit{s} vs \textit{ud}}
        \label{fig:DJT_sVud_eff_p}
    \end{subfigure}
    \hfill
    \begin{subfigure}[t]{0.495\textwidth}
        \centering
        \includegraphics[width=\textwidth]{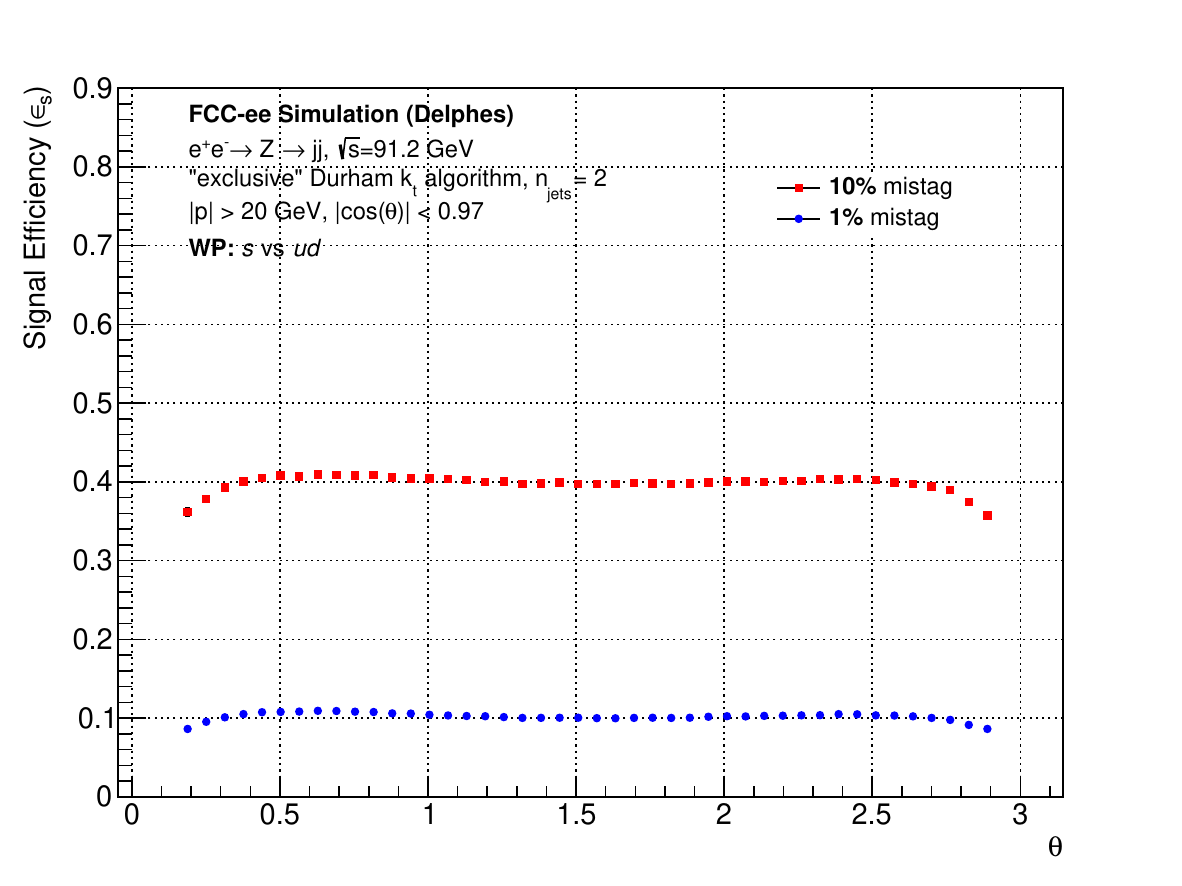}
        \caption{\textit{s} vs \textit{ud}}
        \label{fig:DJT_sVud_eff_theta}
    \end{subfigure}
    \caption{\label{fig:DJT_effs}The jet flavour tagging efficiency over the range of jet momentum and the jet axis polar angle for jets of $e^+e^- \to Z \to q\bar{q}$ events at $\sqrt{s}=91.2$ GeV. Three cases at $1\%$ and $10\%$ background efficiencies are shown: $b$ vs $c$ tagging, $c$ vs $s$ tagging, $s$ vs $ud$ tagging.}
\end{figure}

The tagging efficiency of \texttt{DeepJetTransformer} was evaluated for three cases: $b$ vs $c$ tagging, $c$ vs $s$ tagging, $s$ vs $ud$ tagging. Figure \ref{fig:DJT_effs} shows the efficiency of \texttt{DeepJetTransformer} over the entire jet momentum range and the jet-axis polar angle ($\theta$) range for all three cases for two working points. The efficiency for $b$ vs $c$ tagging and $c$ vs $s$ tagging is mostly uniform, showing a good performance for all jet momenta. Similarly, the performance is largely uniform over the $\theta$ range for all three cases, degrading at the extremes due to jet constituents being lost by fiducial cuts.

However, the $s$ vs $ud$ tagging efficiency displays a peculiar distribution over the momentum range of interest, as shown in Figure \ref{fig:DJT_sVud_eff_p}. This was found to be dependent on the two most distinguishing features for identifying $s$-jets: $K^{\pm}/\pi^{\pm}$ discrimination and V$^0$ reconstruction. The \textit{low-momentum} ($24<|p|<35$ GeV) strange jets, on average, have lower $K^{\pm}$ multiplicities, which leads to a reduced tagging efficiency. The \textit{very-low-momentum} ($|p|<24$ GeV) strange jets have a significantly low total charged-particle multiplicity, making V$^0$ reconstruction crucial. The majority of such jets have a single reconstructed V$^0$, helping identify the $s$-jets. On the other hand, the \textit{low-momentum} strange jets tend to have multiple V$^0$s, splitting the already low jet momentum among these V$^0$s and other hadrons. This is expected to make the strange jet identification more ambiguous. Hence, the $s$-tagging efficiency rises at very low momenta.

A similar but exaggerated trend in the distribution is seen for the looser working point of $10\%$ mistag rate for jets with momentum values below 25 GeV. The efficiency is observed to be stable in momentum above this value. As stated above, some of this increase in $s$-tagging efficiency can be attributed to the presence of a reconstructed V$^0$ in jets with low particle multiplicities. Another important aspect to note is that only a small fraction of jets ($<1\%$) with such very low momenta are present in $Z$ boson decays. This means that these low-momentum jets will not have a large contribution to the training of the neural network or the working point determination, which will both be dominated by the bulk of the momentum distribution. The fact that the $10\%$ $u,~d$-jet background efficiency also increases to $40\%$ for momenta less than 25 GeV implies that this part of the jet momentum phase space is likely not optimally examined by the neural network. A potential method to improve would be to use training weights flattened over the jet momentum and train on much larger samples with this part of the momentum distribution sufficiently populated. But since these jets contribute to a very small fraction of the total $Z$ boson decays, the improvement in analyses requiring strange tagging would likely not be significant unless the physics case is specific.

\subsection{Qualitative Comparison with Other Taggers}
A fair quantitative comparison with other taggers developed for future colliders is not feasible due to differing event samples and input features. However, the jet tagging performance trends are very similar to those of \texttt{ParticleNetIDEA} \cite{Bedeschi_2022, Gautam:2022szi}.  The strange tagging efficiency of \texttt{ParticleNetIDEA} against the $u$-, $d$-jets surpasses that of \texttt{DeepJetTransformer}, owing to PID techniques like cluster counting and time-of-flight used by \texttt{ParticleNetIDEA} and the conservative PID estimates of \texttt{DeepJetTransformer}. A more detailed training dataset including such PID variables is expected to improve the tagging efficiencies of \texttt{DeepJetTransformer}.

\texttt{DeepJetTransformer} outperforms \texttt{ParticleNetIDEA} in bottom-gluon discrimination, especially for efficiencies lower than $90\%$. \texttt{DeepJetTransformer} also has a better discrimination of $b$-jet background for all other signal quark jet flavours. This efficient discrimination can be attributed to the inclusion of SVs. 

With about $10^6$ parameters and efficient transformer blocks as the workhorse, training \texttt{DeepJetTransformer} converges within 2 hours after approximately 50 epochs on an NVIDIA Tesla V100s GPU. The computational complexity, measured in FLOPs, is approximately 19.7 MFLOPs. Comparatively, \texttt{DeepJetTransformer} requires fewer FLOPs than competing architectures \cite{Qu:2019gqs, Qu:2022mxj}, making it an excellent choice to efficiently test the impact of the constantly evolving detector design on flavour tagging.

\subsection{Dependence on the Quality of Particle Identification}
\label{subsec:pid_effect}
Several $K^{\pm}$ classification scenarios were defined by fixing the efficiency of misidentification to $\pi^{\pm}$ and varying the $K^{\pm}$ identification efficiency. In addition, the limiting cases of Kaon identification with $0\%$ and $100\%$ efficiencies were considered. These are referred to henceforth as the no $K^{\pm} \mathrm{ID}$ and the perfect $K^{\pm} \mathrm{ID}$ scenarios. The considered efficiencies and the misidentification rates are the following: 

\begin{table}[h]

    \centering
    \begin{tabular}{c||c | c | c | c | c | c | c | c}
        $K^{\pm}$ ID efficiency & 0\% & 20\% & 40\% & 60\% & 80\% &90\% & 95\% & 100\% \\
        %\hline
        $\pi^{\pm}$ misID efficiency & 0\% & 10\% & 10\% & 10\% & 10\% & 10\% & 10\% & 0\%
    \end{tabular}\caption{\label{tab:PIDsettings}Considered scenarios for $K^{\pm}$ and $\pi^{\pm}$ particle identification performance. }
\end{table}

\begin{figure}[b]
    \centering
    \begin{subfigure}[b]{0.495\textwidth}
        \centering
        \includegraphics[width=\textwidth]{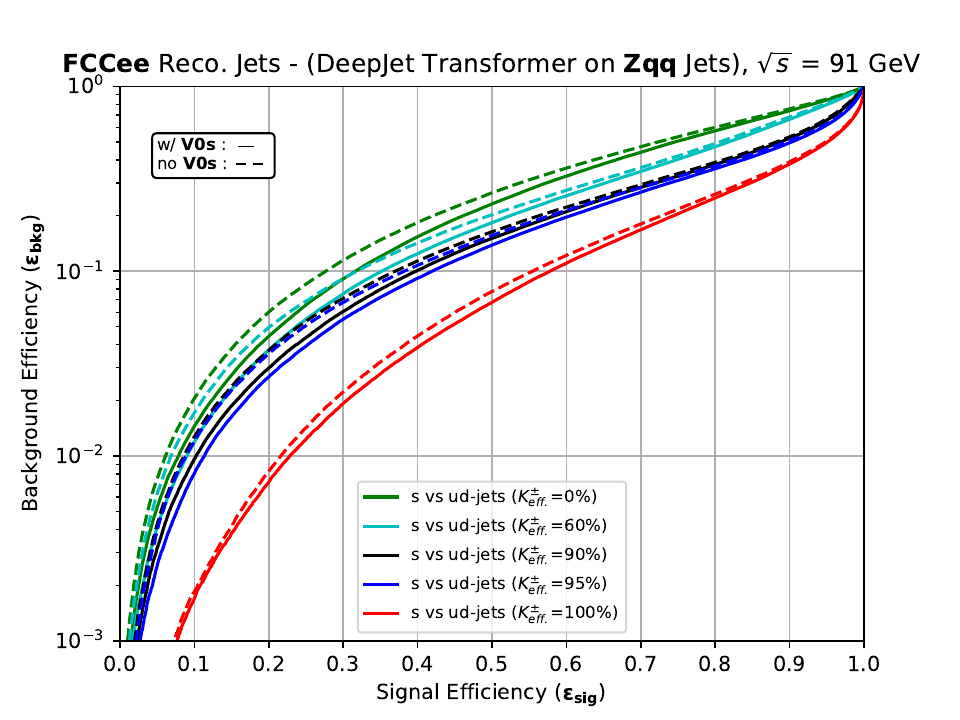}
        \caption{ROC curves}
        \label{fig:stagging_PIDV0_roc}
    \end{subfigure}
    \hfill
    \begin{subfigure}[b]{0.495\textwidth}
        \centering
        \includegraphics[width=\textwidth]{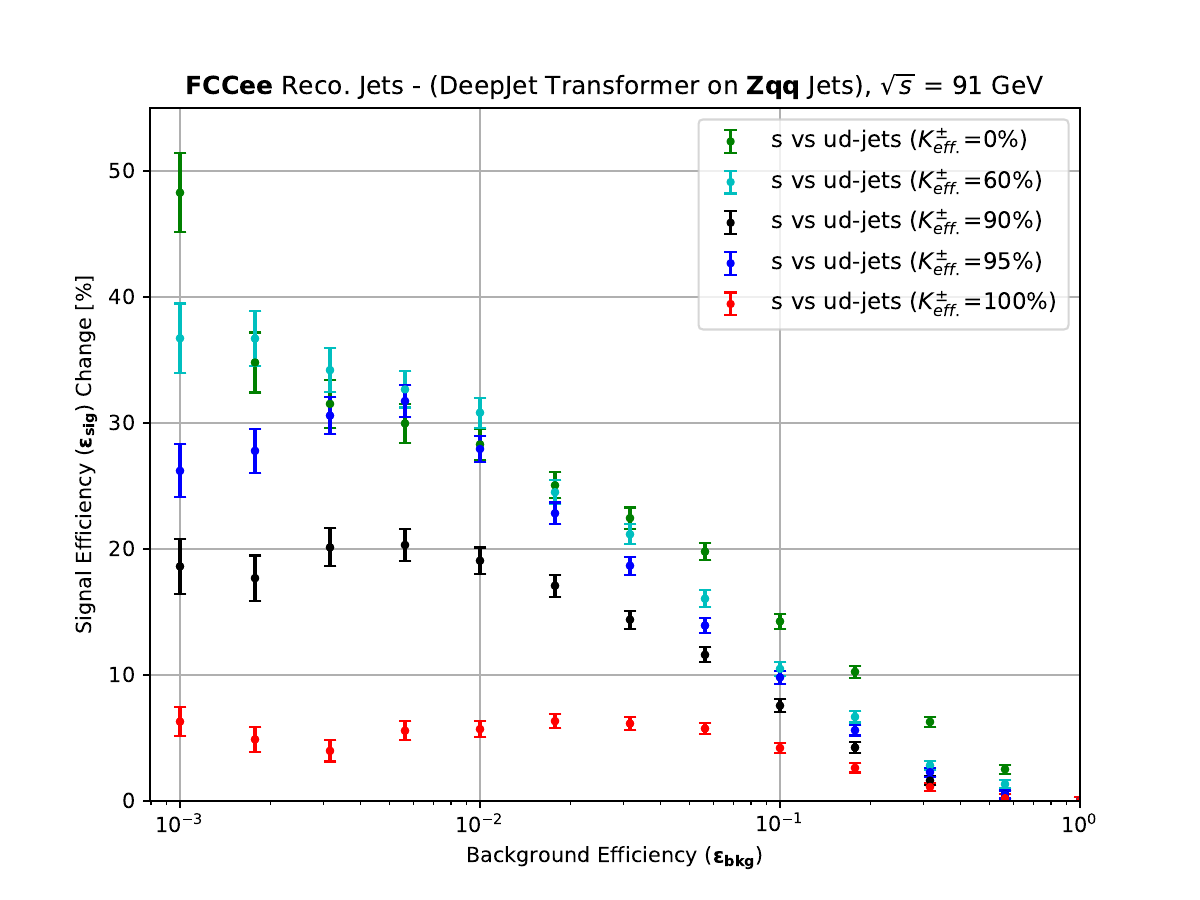}
        \caption{Percent change in signal efficiency}
        \label{fig:stagging_PIDV0_ratio}
    \end{subfigure}
    \caption{The dependence of strange jet tagging performance on the inclusion of V$^{0}$s and charged Kaon identification scenarios. (a) ROC curves for $s$ vs $ud$ tagging at the $Z$ resonance at $\sqrt{s}=91.2$ GeV. Solid lines represent results with the inclusion of V$^0$s, while dashed lines show the results without them. (b) Percent change in signal efficiency ($\epsilon_{sig}$) with the inclusion of V$^0$s for $s$ vs $ud$ tagging for each of the $K^{\pm} \mathrm{ID}$ scenarios listed in Table \ref{tab:PIDsettings}. The axes are swapped with respect to Figure \ref{fig:stagging_PIDV0_roc} to present the percent change in signal efficiency ($\epsilon_{sig}$) as a function of 12 fixed background efficiencies ($\epsilon_{bkg}$).}
    \label{fig:stagging_PIDV0}
\end{figure}

The no $K^{\pm} \mathrm{ID}$ scenario is used as the reference in this section to assess the impact of adding PID variables as input features for jet flavour tagging.
The largest performance gain with the addition of $K^{\pm} \mathrm{ID}$ information is predictably in the classification of $s$ vs $ud$ jets, shown in Figure \ref{fig:stagging_PIDV0}. Relative to the reference no $K^{\pm} \mathrm{ID}$ scenario, with a $\epsilon_{sig}$ of $31.6$\% at a $\epsilon_{bkg}$ of $10$\%, strange tagging efficiency improvements of $11.4\%$, $25.9\%$, and $32.9\%$ are evident as the $K^{\pm} \mathrm{ID}$ efficiency is increased to $60\%$, $90\%$, and $95\%$, respectively. The perfect $K^{\pm} \mathrm{ID}$ scenario shows the most sizeable performance gain in $\epsilon_{sig}$ of $82.9\%$. This large performance improvement over the $95\%$ $K^{\pm} \mathrm{ID}$ efficiency with the efficiency of misidentification to $\pi^{\pm}$ of $10\%$ scenario suggests that minimising this misidentification is crucial to tagging strange jets, given their high $\pi^{\pm}$ multiplicity \cite{Nakai:2020kuu}.

The performance gain for other forms of classification was marginal, with the exception of $c \text{ vs } ud$ and $u \text{ vs } d$ discrimination. For $c \text{ vs } ud$, a performance gain of $1.8 \%$ from a $\epsilon_{sig}$ of $89.3\%$ to $90.9\%$ at a $\epsilon_{bkg}$ of $10$\% is observed while comparing the no $K^{\pm} \mathrm{ID}$ and the perfect $K^{\pm} \mathrm{ID}$ scenarios. In the case of $u \text{ vs } d$, a $12.5\%$ performance gain from a $\epsilon_{sig}$ of $13.6\%$ to $15.3\%$ at a $\epsilon_{bkg}$ of $10$\% is observed.

These results confirm the importance and necessity of particle identification techniques, especially for strange quark studies, as was also noted by some previous studies \cite{Azzi2021,Albert:2022mpk,Bedeschi_2022}.

\subsection{Dependence on the Presence of Neutral Kaons}

As noted earlier, an excess of V$^0$s, reconstructed $K_S^0$ and $\Lambda^0$, carrying the bulk of the jet momenta is also a distinguishing feature of strange jets and these are expected to be more significant in the scarcity of charged Kaons. The inclusion of V$^0$ variables, as Figure \ref{fig:stagging_PIDV0} shows, results in an improvement of signal efficiency ranging from $14.3 \%$ in case of no $K^{\pm} \mathrm{ID}$ to $4.2 \%$ in the case of perfect $K^{\pm} \mathrm{ID}$ at a background efficiency of $10$\% for $s \text{ vs } ud$ discrimination. The percent improvement in signal efficiency for each of the $K^{\pm} \mathrm{ID}$ scenarios listed in Table \ref{tab:PIDsettings} is depicted separately in Figure \ref{fig:stagging_PIDV0_ratio}. This trend proves the importance of V$^0$s to identify strange jets with low $K^{\pm}$ multiplicities or substandard $K^{\pm}/\pi^{\pm}$ discrimination. The performance gain in other forms of classification was again marginal.

\subsection{Importance of Variable Classes and Individual Variables}

Aiming to estimate the relative importance of a given variable class (e.g. SV variables), the classifier performance was evaluated using the Permutation Feature Importance \cite{Breiman2001, fisher2019all} method. 

In particular, the variable class under investigation was shuffled amongst all other jets, keeping the rest of the variables unchanged. Specifically, the values for the variable class under investigation were randomly permuted across all jets in the dataset, while the remaining variables for all jets were left unchanged. This disrupts the relationship between the permuted variable and the jet classification, allowing for an estimate of how much the performance of the classifier depends on the given variable. The resulting performance change was considered for discriminating between $b$- vs $c$-, $c$- vs $s$-, and $s$- vs $ud$- jets, compared to the baseline where no variable classes were permuted. Charged jet constituent variables, listed in Table \ref{tab:jetconstvars}, were found to be the most impactful variable class for all types of discrimination at a background efficiency of $\epsilon_{bkg} = 10\%$, as depicted in Table \ref{tab:fimportglob}. This is presumably due to charged particles being the majority of the reconstructed particles in the jets. SV variables, listed in Table \ref{tab:svvars}, primarily benefited $c$ vs $s$ discrimination, with $s$ vs $ud$ tagging particularly insensitive. Of the remaining three variable classes, V$^0$ variables and neutral jet constituent variables were found to almost exclusively impact the performance of $s$ vs $ud$ discrimination, with little impact on both $b$ vs $c$ and $c$ vs $s$ discrimination, justifying the inclusion of V$^0$s for identifying $s$-jets through conservation of strangeness. Jet-level variables were found to be the least significant, marginally impacting $s$ vs $ud$ discrimination, and having virtually no impact on heavy flavour discrimination. Moving to the high purity regime at a background efficiency of $\epsilon_{bkg} = 0.1\%$, primarily the same trends were observed, with the impact of any variable type being amplified. SV variables, in particular, became hugely important to heavy flavour tagging, reaching almost equal in impact to the charged jet constituent variables, proving that the presence and properties of SVs are definitive indicators for identifying heavy flavour jets.

\begin{table}[h]
\centering
\begin{tabular}{c l|c c c c c} 
\multicolumn{2}{c|}{Variable Class} & Jet-level & Charged & Neutral & SV & V$^{0}$ \\ 
\hline
\hline
& \textit{b} vs \textit{c} & $2.4\%$ & $62.4\%$ & $2.2\%$ & $13.9\%$ & $0.1\%$\\ 
\textbf{$\epsilon_{bkg} = 10\%$} & \textit{c} vs \textit{s} & $1.2\%$ & $65.7\%$ & $2.9\%$ & $29.6\%$ & $0.2\%$\\
& \textit{s} vs \textit{ud} & $7.6\%$ & $59.4\%$ & $21.8\%$ & $5.0\%$ & $16.4\%$\\
\hline
& \textit{b} vs \textit{c} & $6.6\%$ & $97.0\%$  & $8.0\%$ & $89.9\%$ & $0.6\%$\\ 
\textbf{$\epsilon_{bkg} = 0.1\%$} & \textit{c} vs \textit{s} & $9.3\%$ & $96.1\%$ & $11.0\%$ & $77.9\%$ & $0.2\%$\\ 
& \textit{s} vs \textit{ud} & $35.9\%$ & $91.0\%$ & $57.3\%$ & $7.4\%$ & $43.8\%$\\
\end{tabular}
\caption{\label{tab:fimportglob}Performance decrease in signal efficiency ($\epsilon_{sig}$) after permutation of variable classes defined in Section \ref{subsec:input} for fixed background efficiencies ($\epsilon_{bkg}$) of 10\% and 0.1\%.}
\end{table}

The above studies were repeated to estimate the relative importance of individual variables (e.g. $m^{\text{SV}}$), where rather than shuffling an entire variable class amongst jets, one individual variable was shuffled amongst itself. The 64 variables can be loosely split into the following categories:
\begin{itemize}
        \item Kinematic ($|p|$, $E$, $|p|/|p|_{\text{jet}}$, $\theta$, $\Delta\theta$, …)
        \item PID ($isPhoton$, $K^{\pm} \text{ID}$, …)
        \item Track ($D_{0}$, $z_{0}$, …)
\end{itemize}
It was found that, at a background efficiency of $10\%$, kinematic variables of charged particle constituents, including $\dfrac{E_{\text{ch.}}}{E_{\text{jet}}}$ and $\dfrac{|p|_{\text{ch.}}}{|p|_{\text{jet}}}$, were generally impactful, particularly for $c$ vs $s$ discrimination. Track variables, such as $D_{0}/\sigma_{D_{0}}$ and $z_{0}$, were the most impactful, though less for $b$ vs $c$ than other types of discrimination, possibly due to their redundant information after the inclusion of SVs. PID variables had little impact on $b$ vs $c$ and $c$ vs $s$ discrimination, but $K^{\pm} \text{ID}$ and photon ID were the most important for $s$ vs $ud$ discrimination, as was observed earlier. The high purity regime at a background efficiency of $0.1\%$ resulted in similar trends, though with PID variables, including $K^{\pm} \text{ID}$ and photon ID, decreasing in importance and being somewhat replaced by kinematic ones. It should be stated that the baseline $K^{\pm} \text{ID}$ scenario, as mentioned in Section \ref{sec:performance}, is deliberately pessimistic, which could account for its decrease in importance. Track variables remained the most impactful. The secondary vertex mass $m^{\text{SV}}$ became the most impactful variable in $b$ vs $c$ discrimination at high purity by a sizeable margin, as SV kinematics store essential information about the decaying hadrons. The results of this study are summarised in Table \ref{tab:fimportloc} below.

\begin{table}[h]
\centering
\begin{tabular}{c l|c c c c c c c} 
\multicolumn{2}{c|}{Variable} & $\ln(E_{\text{ch.}})$ & \texttt{isPhoton} & $K^{\pm} \text{ID}$ & $m^{\text{SV}}$ & $|p|^{\text{V}^{0}}$ & $z_{0}$ & $D_{0}/\sigma_{D_{0}}$\\ 
\hline
\hline
& \textit{b} vs \textit{c} & $3.5\%$ & $0.3\%$ & $0.2\%$ & $3.0\%$ & $0.1\%$ & $7.8\%$ & $11.6\%$\\ 
\textbf{$\epsilon_{bkg} = 10\%$} & \textit{c} vs \textit{s} & $23.8\%$ & $0.7\%$ & $0.5\%$ & $0.3\%$ & $0.2\%$ & $20.9\%$ & $39.1\%$\\
& \textit{s} vs \textit{ud} & $12.8\%$ & $16.6\%$ & $38.8\%$ & $0.0\%$ & $9.2\%$ & $23.3\%$ & $26.7\%$\\
\hline
& \textit{b} vs \textit{c} & $13.8\%$ & $1.3\%$ & $0.9\%$ & $67.2\%$ & $0.8\%$ & $34.1\%$ & $45.0\%$\\ 
\textbf{$\epsilon_{bkg} = 0.1\%$} & \textit{c} vs \textit{s} & $57.6\%$ & $0.9\%$ & $4.8\%$ & $7.0\%$ & $0.3\%$ & $56.2\%$ & $79.5\%$\\ 
& \textit{s} vs \textit{ud} & $35.0\%$ & $28.0\%$ & $59.0\%$ & $0.4\%$ & $34.7\%$ & $60.5\%$ & $80.1\%$\\
\end{tabular}
\caption{\label{tab:fimportloc}Performance decrease in signal efficiency ($\epsilon_{sig}$) after permutation of individual variables defined in Section \ref{subsec:input} for fixed background efficiencies ($\epsilon_{bkg}$) of 10\% and 0.1\%. A set of seven variables, chosen among the most impactful, is presented here.}
\end{table}

\subsection{Dependence on the Flavour Definition}
Defining the flavour of a reconstructed jet is a complex task. Several definitions have been used in past and current experiments to assign the flavour of MC-generated jets. The flavour definition can impact the classifier performance for $Z$ boson decay events because this definition can lead to jets being assigned a different flavour than the original quark.

In the $Z$ boson definition used throughout this work as introduced in Section \ref{subsec:sim}, the flavour of a jet is defined as the flavour of the quark to which the $Z$ boson decays. The hadronisation and fragmentation of the quark are ignored in this definition. One flavour definition that accounts for fragmentation and hadronisation effects is the Ghost Matching algorithm used at CMS\ \cite{CACCIARI2008119}, which defines the flavour of a jet by finding the hadrons or partons from the MC history of the jet, clustered with the same jet clustering algorithm as the reconstructed jet.

The largest performance differences of the algorithm after changing flavour definitions can be observed in the discrimination of $s$-jets vs $ud$-jets, where the Ghost Matching definition leads to a 11.8\% higher tagging efficiency than the $Z$ boson definition at a fixed background efficiency of 10\%. Such significant changes in performance make it essential to account for the used flavour definitions while comparing different flavour tagging algorithms.

\section{Example of Performance: The \textit{Z} Boson at the FCC-ee} \label{sec:Zss}

The $Z$ boson decays relatively uniformly to the five quark flavours, and none of the decay channels to $q\bar{q}$ pairs are suppressed. Thus, tagging a particular jet flavour entails discrimination against every other flavour. Especially, isolating $Z \to s\bar{s}$ events from the exclusive decays of the $Z$ boson provides a challenging case to tag the $s$-jets by eliminating both the heavy jets and $u$-, $d$-jets. The dominant discriminating variable against the heavy jets is the reconstructed SVs, while it is the presence of a high-momentum strange hadron against $u$- and $d$-jets. This makes isolating $Z \to s\bar{s}$ events from the exclusive hadronic decays of the $Z$ boson an ideal metric to assess the performance of \texttt{DeepJetTransformer} in the FCC-ee environment and allows for a unique opportunity to access a hitherto scarcely studied channel. Further backgrounds are not considered but are expected to be well under one per cent of the total expected yield around the $Z$ boson resonance.

\subsection{Event and Jet Selection}

The $e^{+}e^{-} \rightarrow Z \rightarrow q\bar{q}$ event samples with $q \equiv b, c, (u, d, s)$ described in Section \ref{subsec:sim} are used. These are the same samples used to evaluate the performance of \texttt{DeepJetTransformer}.

Events are selected if exactly two jets could be reconstructed with their final constituents. Jets with low momentum or jet axes outside the fiducial region of the detector are excluded. An event is selected if both of its jets have a 3-momentum magnitude ($|p|$) greater than $20$ GeV and the polar angle ($\theta$) of their jet axes within $14$ and $176$ degrees. Events are required to have jets of the same MC flavour, defined as the flavour of the quarks to which the $Z$ boson decays.

\subsection{Performance and Working Points}

All jets from $Z\to q\bar{q}$ events are independently evaluated using \texttt{DeepJetTransformer}. Discriminants are defined to sequentially remove the heavy flavour background ($b$- and $c$-jets) and the light flavour background ($u$- and $d$-jets). The $s$-jets are first tagged to be discriminated from $b$- and $c$-jets by defining the discriminant as in Eq. \ref{eq:discriminant} with $s$-jets as signal and $b$- and $c$-jets as background. For the jets tagged by introducing a cut on this discriminant, another discriminant is defined to distinguish $s$-jets from $u$- and $d$-jets through the same method. The signal efficiencies after each subsequent cut, corresponding to four working points with increasing purity, are reported in Table \ref{tab:cutflow}.

\begin{table}[ht]
\centering
\begin{tabular}{c l|c c c c} 
& & Mistag Rate [\%] & Efficiency [\%] & $N_{sig}$ & $N_{bkg}$ \\ 
\hline
\hline
\textbf{WP1} & \textit{s} vs \textit{bc} & $10$ & $98.93 \pm 0.03$ & $7.35\times10^{11}$ & $1.35\times10^{12}$ \\ 
& \textit{s} vs \textit{ud} & $10$ & $40.03 \pm 0.04$ & $1.45\times10^{11}$ & $3.25\times10^{10}$ \\
\hline
\textbf{WP2} & \textit{s} vs \textit{bc} & $1$ & $54.18 \pm 0.04$  & $2.38\times10^{11}$ & $2.06\times10^{11}$ \\ 
& \textit{s} vs \textit{ud} & $10$ & $39.28 \pm 0.06$ & $5.10\times10^{10}$ & $5.57\times10^{9}$ \\ 
\hline
\textbf{WP3} & \textit{s} vs \textit{bc} & $1$ & $54.18 \pm 0.04$ & $2.38\times10^{11}$ & $2.06\times10^{11}$ \\ 
& \textit{s} vs \textit{ud} & $1$ & $10.05 \pm 0.11$ & $1.12\times10^{10}$ & $4.77\times10^{8}$ \\ 
\hline
\textbf{WP4} & \textit{s} vs \textit{bc} & $0.1$ & $17.96 \pm 0.06$ & $3.23\times10^{10}$ & $6.98\times10^{9}$ \\ %
& \textit{s} vs \textit{ud} & $0.1$ & $1.98 \pm 0.33$  & $3.56\times10^{8}$ & $3.38\times10^{6}$ \\ 

\end{tabular}
\caption{\label{tab:cutflow}Presented are the efficiencies to select $s$ quark jets and the mistag rate for other flavours at four different working points. Also listed are the expected yields calculated for an integrated luminosity of $125~\text{ab}^{-1}$. Signal is defined as $Z \to s\bar{s}$ events while the background is composed of $Z \to q\bar{q}$ (all quarks but $s$ quarks) events. The number of observed events is significantly above the canonical discovery significance of five standard deviations for all selections.}
\end{table}

\begin{figure}%[hptb]
    \centering
    \begin{subfigure}[t]{0.495\textwidth}
        \centering
        \includegraphics[width=\textwidth]{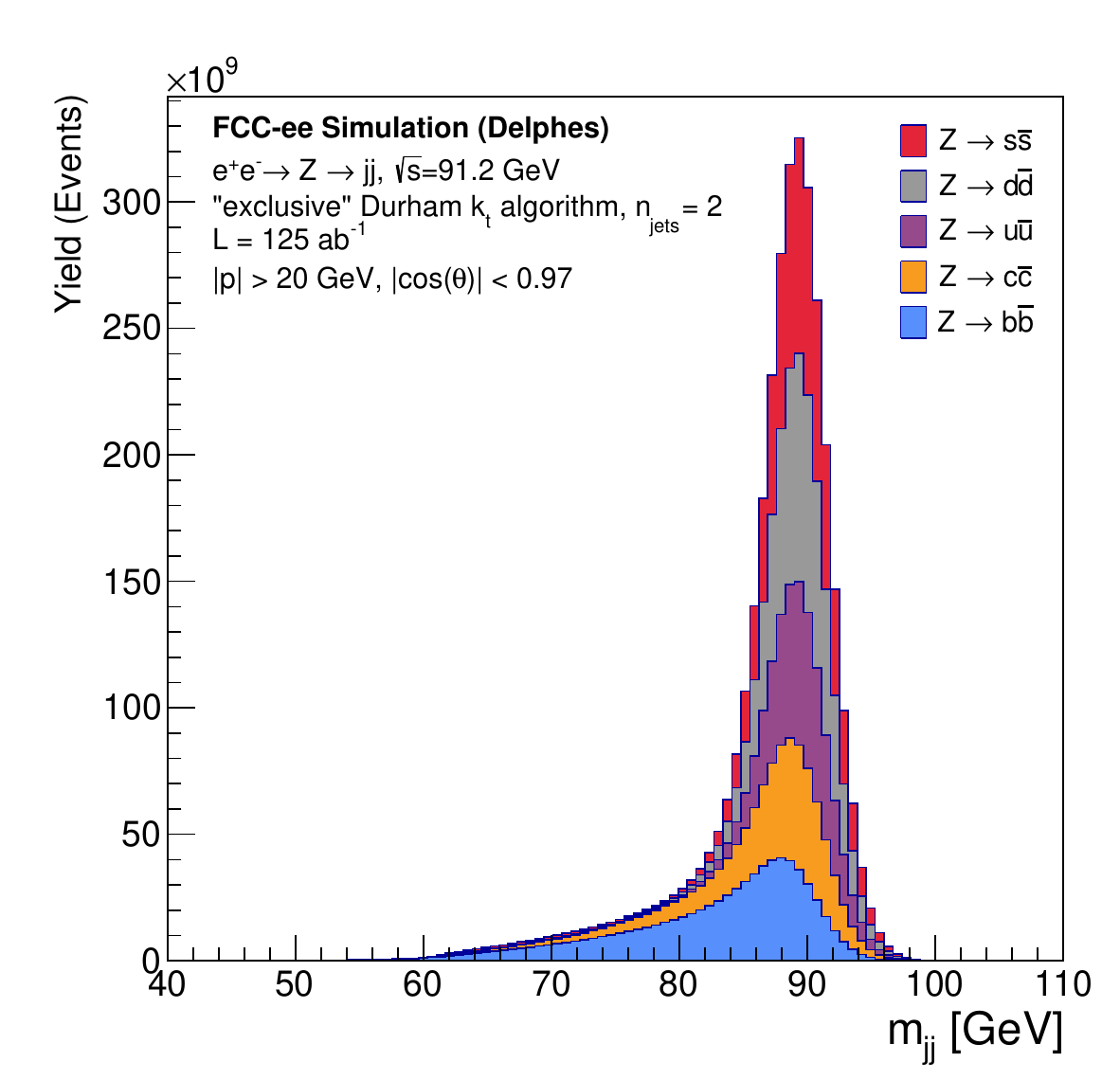}
        \caption{}
        \label{fig:Z_beforetag}
    \end{subfigure}
    \hfill
    \begin{subfigure}[t]{0.495\textwidth}
        \centering
        \includegraphics[width=\textwidth]{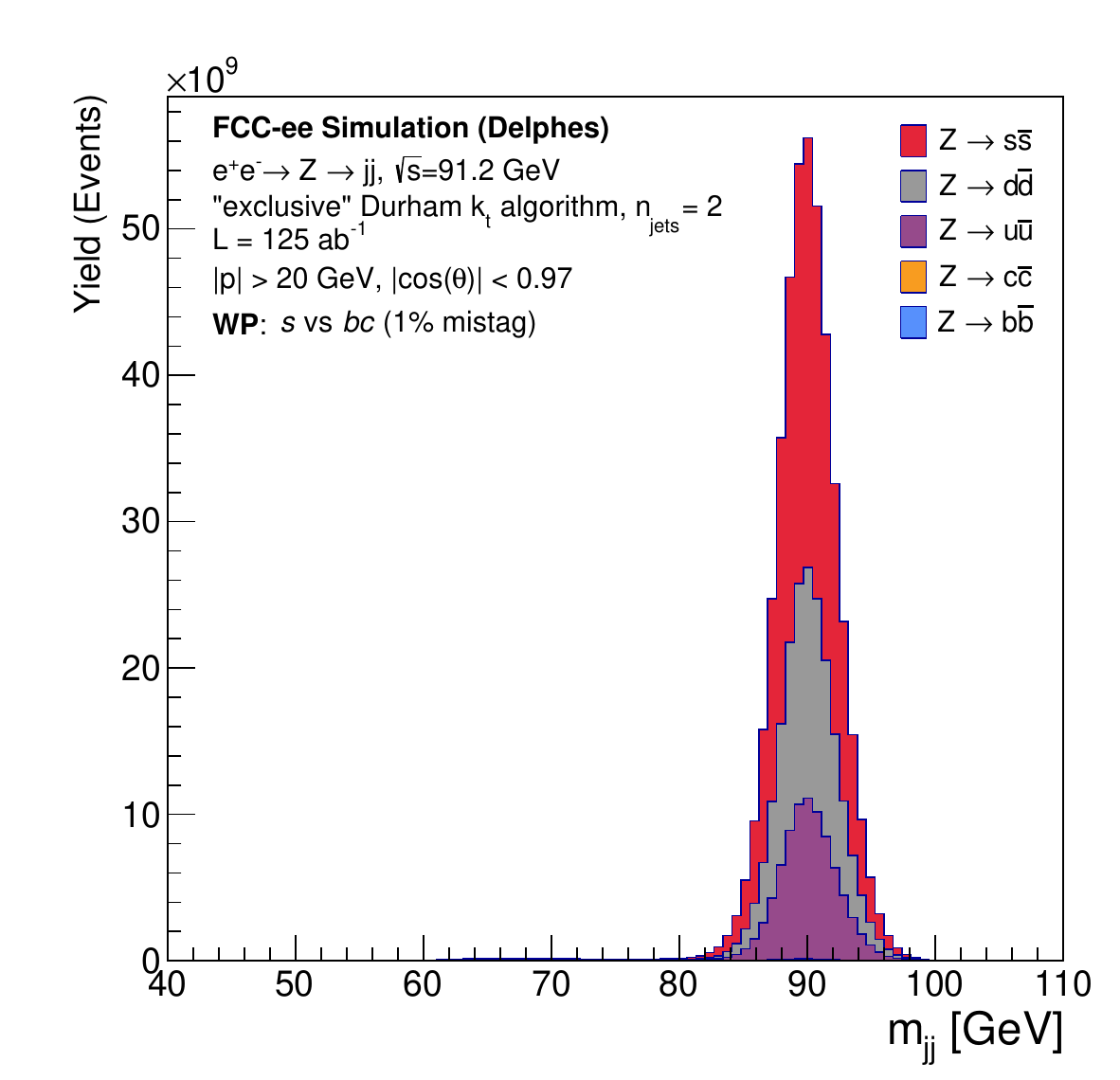}
        \caption{}
        \label{fig:Z_aftertag_sbc}
    \end{subfigure}
    \hfill
    \begin{subfigure}[t]{0.495\textwidth}
        \centering
        \includegraphics[width=\textwidth]{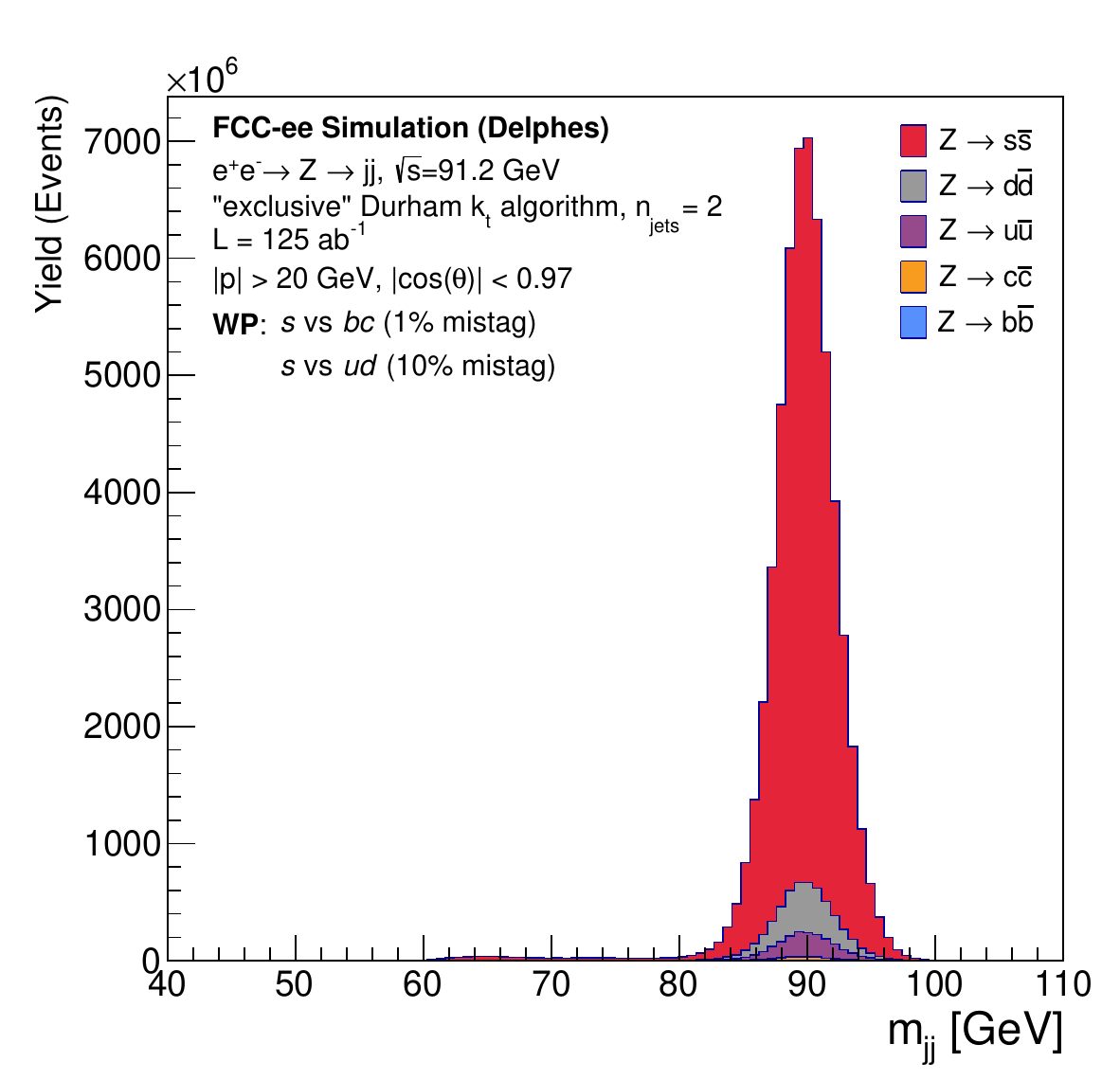}
        \caption{}
        \label{fig:Z_aftertag_sud_10}
    \end{subfigure}
    \hfill
    \begin{subfigure}[t]{0.495\textwidth}
        \centering
        \includegraphics[width=\textwidth]{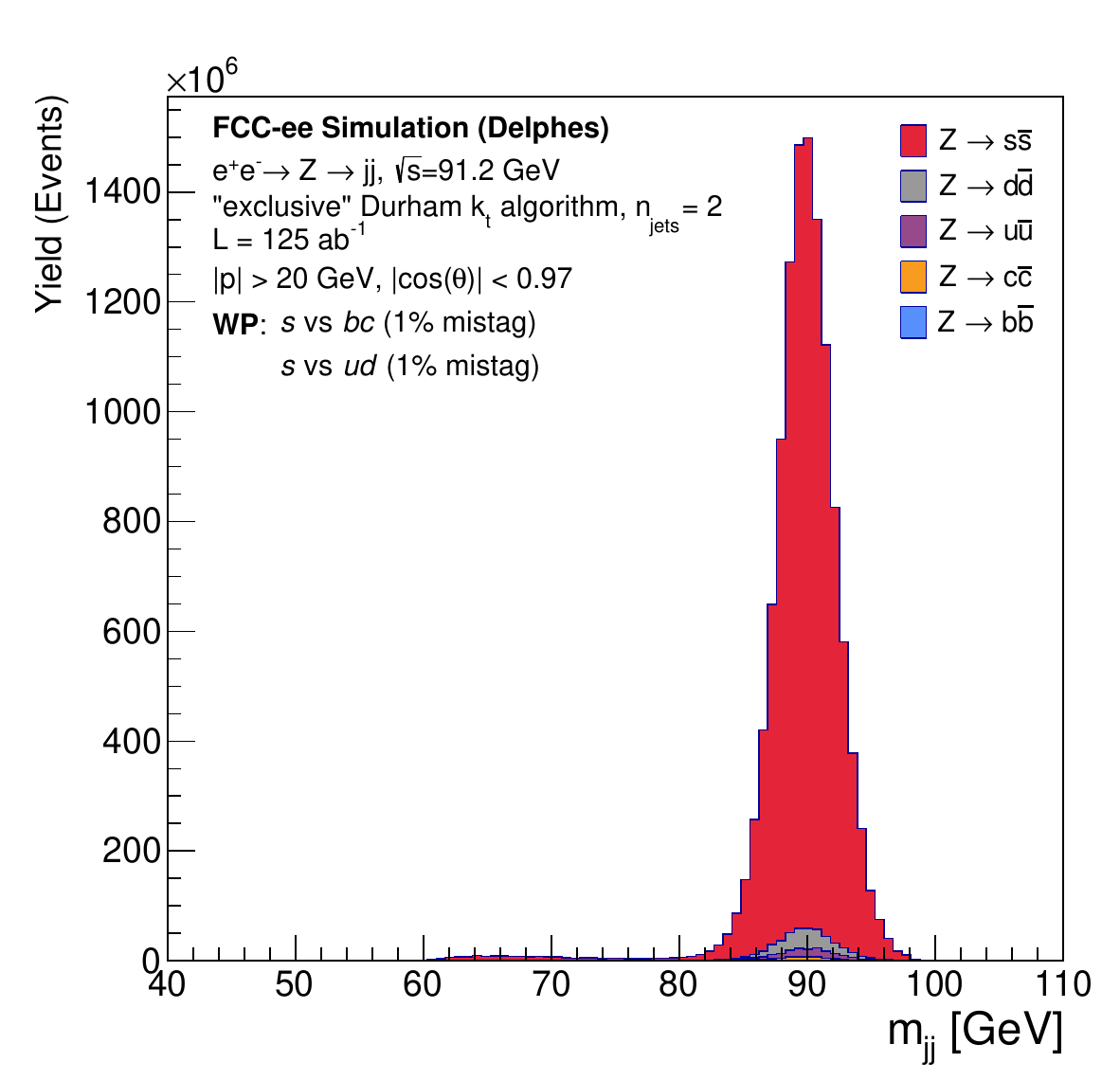}
        \caption{}
        \label{fig:Z_aftertag_sud_01}
    \end{subfigure}
    \caption{\label{fig:Zpeak_tagvsnotag}The reconstructed invariant mass of the dijet system before and after tagging both jets with \texttt{DeepJetTransformer}, corresponding to WP2 and WP3 in Table\ \ref{tab:cutflow}, for an assumed integrated luminosity of $125~\mathrm{ab}^{-1}$. Both jets are required to be tagged in each case. Shown are (a) the distribution without tagging applied, (b) after the rejection of $b$- and $c$-jets vs $s$-jets at $1\%$ mistag rate, (c) the distribution after rejection of $b$- and $c$-jets at $1\%$ and $u$- and $d$-jets vs s-jets at $10\%$ mistag rate, (d) the distribution after rejection of  $b$- and $c$-jets at $1\%$ and $u$- and $d$-jets vs s-jets at $1\%$ mistag rate.}
    \label{fig:Zss}
\end{figure}

The $Z$ boson resonance is reconstructed from the 4-momentum of the two jets. The reconstructed invariant dijet mass distribution, separated by the MC flavour of the resulting hadronic jets, is shown in Figure \ref{fig:Z_beforetag}. The hadrons in $b$-jets tend to have longer decay chains, which causes more momentum to be lost via neutrinos, resulting in a wider invariant mass distribution for $Z \to b\bar{b}$. Similarly, the $Z \to c\bar{c}$ reconstructed invariant mass distribution also shows a tail, but for the lighter flavour jets, $s$, $u$, and $d$, a clear Gaussian peak can be seen at the $Z$ resonance.

These jets are first tagged to remove the background of $b$- and $c$-jets by defining the discriminant, as described above. If both jets from a $Z$ boson decay event pass this tagging requirement, they are used to reconstruct the invariant mass. The distribution of this invariant mass is displayed in Figure \ref{fig:Z_aftertag_sbc}, with the contributions of the MC flavours of the jets indicated. The jets from the events passing the anti-$b/c$ tag requirement are subsequently tagged with the $s$ vs $ud$ quark tagger to remove the background of $u$- and $d$-jets. Figures \ref{fig:Z_aftertag_sud_10} and \ref{fig:Z_aftertag_sud_01} show the distribution of the reconstructed invariant mass of the $Z$ boson after this additional tag. Both jets of every event are required to pass the tagging requirements in each stage of the selection.

\begin{figure}
    \centering
    \includegraphics[width=\textwidth]{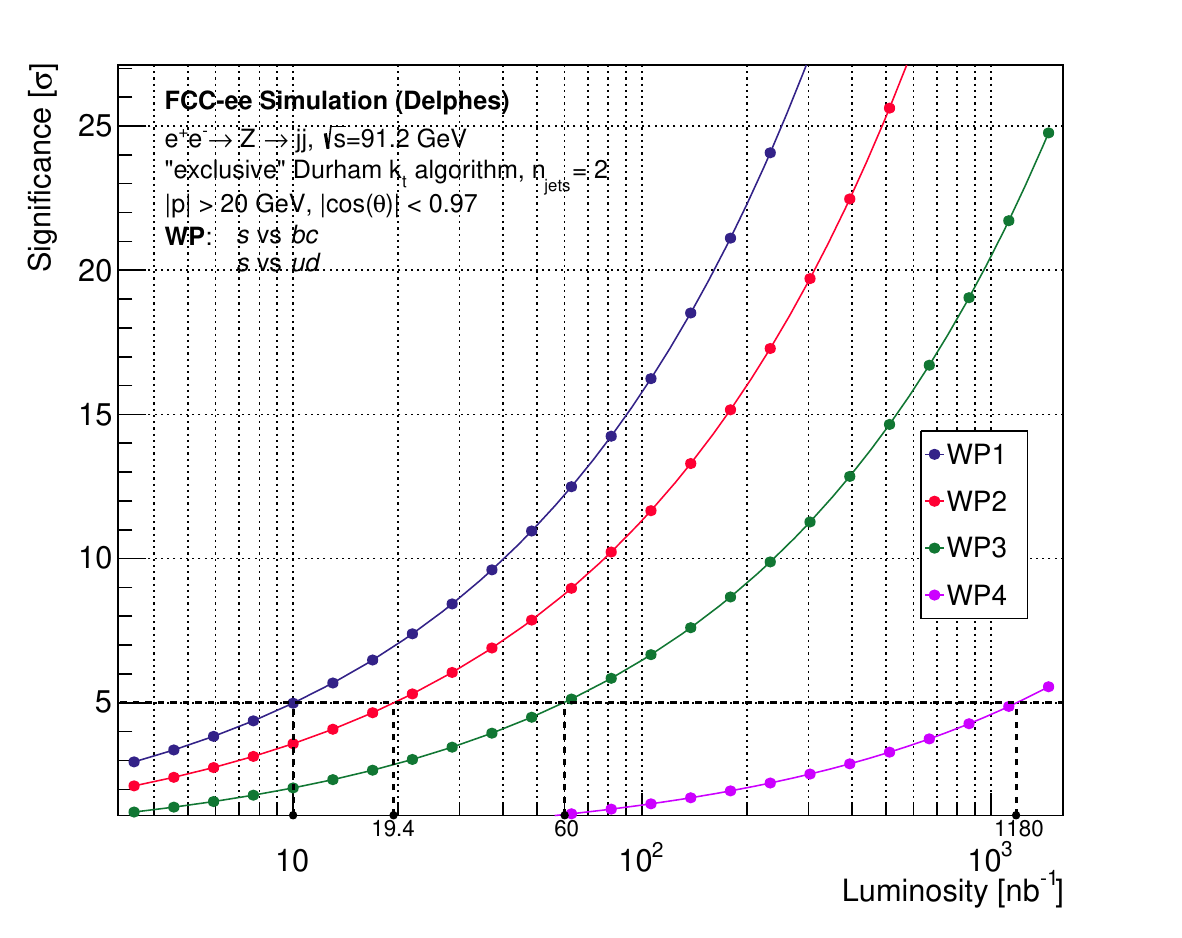}
    \caption{Discovery Significance vs Luminosity for the four working points corresponding to Table \ref{tab:cutflow}. The points noted by the $x$ axis intersecting with the dashed vertical lines are the luminosities required at the four respective working points to achieve the canonical discovery significance of $5\sigma$.}
    \label{fig:disc_sig}
\end{figure}

The reconstructed tagged $Z$ resonance in Figure \ref{fig:Zss} shows that the $Z \to s\bar{s}$ sample is extremely pure after requiring the two consecutive tags on each jet from $Z$ boson decay events. Table \ref{tab:cutflow} lists events corresponding to an integrated luminosity of $125~\mathrm{ab}^{-1}$. The discovery significance, $Z$, in $\sigma$, is defined \cite{sig_calc_note} as,
\begin{equation}
    Z = \sqrt{2\left[ (N_{sig} + N_{bkg}) \log\left(1+\dfrac{N_{sig}}{N_{bkg}}\right) - N_{sig} \right]}.
\label{eq:significance}
\end{equation}
$N_{sig}$ and $N_{bkg}$ refer to the number of signal and background events, respectively. Signal is defined as $Z \to s\bar{s}$ events while the background is composed of $Z \to q\bar{q}$ (all quarks but $s$ quarks) events. It is apparent that all four working points are significantly above the canonical discovery significance of $5\sigma$. It is important to realise that machine backgrounds and irreducible backgrounds from other standard model processes are not considered in this study, and are at the per cent level. However, the remarkable sensitivity warrants investigation of how limited the integrated luminosity can be to observe $Z \to s\bar{s}$ in the considered scenario.

Figure \ref{fig:disc_sig} shows the discovery significance of the process $Z \to s\bar{s}$, under the background-free scenario, as a function of integrated luminosity. The corresponding values of $N_{sig}$ and $N_{bkg}$ at each working point can be referred to from Table \ref{tab:cutflow}. It can be seen that a $5\sigma$ significance can be achieved with minuscule luminosities compared to the FCC-ee run plan, even at the tightest working point. For WP3, corresponding to Figure \ref{fig:Z_aftertag_sud_01}, a $5\sigma$ significance can be reached with a luminosity of $60~\text{nb}^{-1}$, equivalent to {\em less than a second} of the FCC-ee run at the $Z$ resonance.

Data-to-simulation scale factors for $b$-jets can be measured with a precision of approximately $\pm2.5\%$ for jets with $30<p_{\text{T}}<50$ GeV at the LHC experiments. Tagging algorithms at the future colliders are expected to achieve smaller uncertainties.

These findings will open up avenues at FCC-ee for measurements that require ultra-pure $Z \to q\bar{q}$ samples, at least for the three heaviest flavours to which the $Z$ boson decays. Some examples are vector and axial couplings of the $Z$ to up- and down-type quarks and possibly even individual quark flavours and asymmetry parameters of the $Z$ boson in the hadronic decay channels. LEP and SLD performed comprehensive measurements of the forward-backwards charge asymmetry for $e^+e^- \to b\bar{b}$ \cite{Workman:2022ynf}, similar precise measurements for the charm and the strange quark, and possibly the $u$, $d$ quarks, will become feasible at the FCC-ee.

\section{Summary} \label{sec:summary}
The transformer-based model presented in this work can be trained considerably more quickly than the state-of-the-art graph neural network-based taggers \cite{Qu:2022mxj, 10363595}. The discrimination power of this framework called \texttt{DeepJetTransformer} is presented for FCC-ee, allowing the classification of all jet flavours in $e^+e^-$ collisions at the $Z$ resonance.

A tagging efficiency for $b$-jets of about $99\%$($86\%$) can be achieved against $s$-, $u$-, and $d$-jets (c-jets) at a background efficiency of $0.1\%$, pointing to an excellent $b$-jet discrimination, dominantly owing to the secondary vertex reconstruction coming from the expected excellent detector resolution. A $c$-jet tagging efficiency of about $90\%$($70\%$) can be achieved when discriminating from $b$-jets, at a background efficiency of $10\%$($1\%$).

Excellent discrimination can be achieved for $s$-quark tagging against the $b$- and $c$-quark jet background. Against the most challenging background of $ud$-jets, a $40\%$ efficiency for $s$ quark jets can be achieved at a background efficiency of $10\%$. This performance is partially attributed to the inclusion of V$^0$s. Further significant performance enhancement in strange tagging is seen when $K^{\pm}/\pi^{\pm}$ discrimination is included. Minor discrimination can even be achieved between $u$- and $d$-jets. The $Z \to s\bar{s}$ process can be efficiently isolated from other hadronic decays of the $Z$ boson, and an extremely pure $Z \to s\bar{s}$ sample can be obtained.

\section{Outlook} \label{sec:outlook}

The current input feature set is likely far from optimal and could be extended to incorporate further parameters, including those related to jet-shape variables or the full covariance matrix. A primary focus would be to include more realistic PID assumptions based on a specific detector scenario. In Ref. \cite{Bedeschi_2022}, for instance, the mass calculated from the time-of-flight ($m_{t.o.f.}$) and the number of primary ionisation clusters along the track ($dN/dx$) are directly fed as inputs to the NN. On the other hand, it is also evident from the feature importance studies that there is some overlap in the current feature set, which could likely be reduced with marginal impact on the discriminative performance, thus lowering computational complexity if paired with a simplified architecture.

There is also significant room for hyperparameter tuning. The used batch size of 4000 is comparatively large, with typical values being less than 1024. The large batch size was chosen for training stability but has been shown to potentially lead to poorer generalisation. The chosen number of training jets of $O(10^{6})$ can be considered a rough lower bound given the number of parameters in the network $\sim 10^{6}$. A natural next step would be to train the network on a much larger number of jets. Further improvements in the network architecture, including adjustments to layer parameters and network structure, are likely possible, though this was not explored in the context of these studies.

Subdividing jet flavours into categories with unique signatures, such as $b$-jets into those that decay hadronically and semi-leptonically, or $g \to b\bar{b}$ splittings that do not resemble the typical radiation pattern of a gluon jet, is likely to improve discrimination performance. Additional categories could likewise be included for anti-quarks, which would be helpful in discriminating dijet events where a quark-antiquark pair is expected, such as in $Z \to s\bar{s}$ decays. More generally, much could be gained from event-level tagging, particularly for $s$ quark jets, where discrimination comes primarily from a high-momentum Kaon. Tagging an entire event could require not only a high-momentum Kaon in one jet but also an oppositely-charged high-momentum Kaon in the other, thus discriminating against Kaons produced during the dressing of a $u$ or $d$ quark, which will not have an oppositely-charged high-momentum Kaon in the other hemisphere. 

The updated design of the IDEA detector concept has the innermost layer of the vertex detector at 1.2 mm instead of 1.7 mm. It will improve the impact parameter resolution and, consequently, the displaced vertex resolutions, thus enhancing the performance of heavy flavour tagging. Further improvement is expected from an ultra-light ALICE ITS3-like vertex detector \cite{Freitag:2851362}. An updated version of CLD \cite{bacchetta2019cld} is being developed with a dedicated compact-RICH PID detector, ARC, which is expected to aid in strange tagging.

A natural extension of isolating $Z \to s\bar{s}$ events would be to measure the branching fraction and coupling of the $Z$ boson to the $s$ quark and assess further flavour-dependent properties at the $Z$ pole that are sensitive to extensions of the standard model. Extrapolating the excellent performance of \texttt{DeepJetTransformer} in discriminating strange jets and the continuing improvement of jet flavour taggers along with more sophisticated inputs, there is clear potential for the precise study of the light $u$ and $d$ quarks at the $Z$ resonance at the FCC-ee.

The similar performance in Higgsstrahlung events suggests the opportunity to measure the Yukawa coupling of the $s$ quark, as attempted in Ref. \cite{DuarteCampderros2020,Albert:2022mpk}, and the decent gluon discrimination, especially against heavy quarks, will make gluon final states accessible as well. The much larger $Z$ boson cross-section will also provide opportunities for calibration and performance validation on data before the Higgs boson decay to $s$ quarks is examined, which is likely to reduce experimental uncertainties.

\section{Conclusion} \label{sec:conclusion}
Deep learning techniques have demonstrated excellent performance in analysing complex jet structures and extracting subtle flavour signatures in jet flavour identification. The short training time of \texttt{DeepJetTransformer} makes it uniquely suited for prospective studies of the developing detector concepts. It should be noted that even though this study focuses on FCC-ee and the IDEA detector, the conclusions are general, and \texttt{DeepJetTransformer} can also be utilised at other collider projects with appropriate adjustments, such as tuning to different detector geometries, jet clustering algorithms, or energy regimes.
 
These results show that modern jet flavour tagging techniques can isolate very pure samples of $s$ quark decays originating from vector bosons. 
We hope that strange jet tagging will create opportunities for a new category of potential studies at future lepton colliders, including assessment of the feasibility of completely new or more precise measurements and enhancement of the sensitivity to new physics phenomena.

\begin{center} \textbf{Acknowledgments} \end{center}
We want to thank our CMS colleagues at the IIHE in Brussels, especially A.R. Sahasransu and Lode Vanhecke for their preparatory work, and Emil Bols for valuable discussions regarding \texttt{DeepJetTransformer}. We would also like to thank Kyle Cormier at the UZH for helpful discussions regarding the feature importance studies. We are grateful to Frank Gaede, Loukas Gouskos, and Michele Selvaggi for their feedback on the manuscript.

This project is supported by the European Union's Horizon 2020 research and innovation programme under grant agreement No 951754. Kunal Gautam and Eduardo Ploerer are supported by FWO (Belgium) and SNF (Switzerland). Freya Blekman acknowledges support from DESY (Hamburg, Germany), a member of the Helmholtz Association HGF, and support by the Deutsche Forschungsgemeinschaft (DFG, German Research Foundation) under Germany’s Excellence Strategy -- EXC 2121 "Quantum Universe" -- 390833306. Armin Ilg is supported by SNF in Switzerland.

\begin{center} \textbf{Author Contribution} \end{center}
 KG implemented and AI tested the vertex reconstruction algorithm. ADM designed and implemented the \texttt{DeepJetTransformer} architecture. EP adopted the model for the FCC-ee environment, performed input feature selection, and trained and evaluated the performance for several scenarios.
 KG assessed the classifier performance by isolating $Z \rightarrow s\bar{s}$ events from the exclusive $Z \rightarrow q\bar{q}$ decays. FB and AI supervised and reviewed the work throughout the study. FB, ADM, KG, AI, and EP contributed to the writing and editing of this paper.

%
%%%%%%%%%%%%%%%%%%%%%%%%%%%%%%%%%%%%%%%%%%%%%%%%%%%%%%%%%

\newpage
\bibliographystyle{JHEP}
\bibliography{main}

%%%%%%%%%%%%%%%%%%%%%%%%%%%%%%%%%%%%%%%%%%%%%%%%%%%%%%%%%

\end{document}